\documentclass[english,aps,prl,twocolumn, superscriptaddress,showpacs, amsmath]{revtex4-1}

\usepackage[T1]{fontenc}
\usepackage[latin9]{inputenc}
\usepackage{color}
\usepackage{units}
\usepackage{amssymb}
\usepackage{graphicx}
\usepackage{esint}
\usepackage{bm}
\usepackage{natbib}
\usepackage[colorlinks=true, citecolor=blue]{hyperref}
\setcitestyle{journalcolor= blue}

\usepackage{babel}

\begin{document}
\title{Probing 2D Black Phosphorus by Quantum Capacitance Measurements}

\author{Manabendra Kuiri}
\altaffiliation{Contributed equally to this work}
\author{Chandan Kumar}
\altaffiliation{Contributed equally to this work}
\author{Biswanath Chakraborty}
\author{Satyendra N Gupta}
\author{Mit H.Naik}
\author{Manish Jain}
\author{A.K. Sood}
\author{Anindya Das}
\email{anindya@physics.iisc.ernet.in}
\affiliation{Department of Physics, Indian Institute of Science, Bangalore 560012, India}

\begin{abstract}
\textbf{Abstract: }Two-dimensional materials and their heterostructures have emerged as a new class of materials for not only fundamental physics but also for electronic and optoelectronic applications. Black phosphorus (BP) is a relatively new addition to this class of materials. Its strong in plane anisotropy makes BP a unique material to make conceptually new type of electronic devices. However, the global density of states (DOS) of BP in device geometry has not been measured experimentally. Here we report the quantum capacitance measurements together with conductance measurements on a hBN protected few layer BP ($\sim$ 6 layer) in a dual gated field effect transistor (FET) geometry. The measured DOS from our quantum capacitance is compared with the density functional theory (DFT). Our results reveal that the transport gap for quantum capacitance is smaller than that in conductance measurements due to the presence of localized states near the band edge. The presence of localized states is confirmed by the variable range hopping seen in our temperature-dependence conductivity. A large asymmetry is observed between the electron and hole side. The asymmetric nature is attributed to the anisotropic band dispersion of BP. Our measurements establish the uniqueness of quantum capacitance in probing the localized states near the band edge, hitherto not seen in the conductance measurements.
\end{abstract}

\maketitle
\section{Introduction}

Black Phosphorus (BP) is a recent addition to the family of 2D materials like graphene, hBN and layered transition metal dichalcogenides (TMDs).~\cite{ling2015renaissance,li2014black}
BP has a layer dependent direct band gap at $\Gamma$-point ($\sim$ 1.7 eV for a monolayer and $\sim$ 0.3 eV for the bulk),~\cite{PhysRevB.89.235319,qiao2014high,wang2015highly,PhysRevB.90.075434,castellanos2014isolation,das2014tunable}
lying in between the two extremes of other 2D materials, namely semimetallic graphene (no gap) and insulating hBN. The unique aspect of BP is the puckered honeycomb structure of the layer, having two kinds of P-P bonds: the shorter bond length (2.224 \AA) connecting nearest P atoms in the same plane and longer bond length (2.244 \AA) connecting P atoms between the bottom and top of a single layer. Further, the edges are zigzag and armchair along two orthogonal directions, resulting in anisotropic electronic transport ~\cite{qiao2014high,xia2014rediscovering,liu2014phosphorene,yuan2015transport} and optical absorption.~\cite{qiao2014high,wang2015highly,PhysRevB.89.235319} The effective mass of electrons along the zigzag is $\sim$ 10 times higher compared with that along the  armchair direction.~\cite{ling2015renaissance}
 This in-plane anisotropy makes it unique and an attractive choice for electronic and opto-electronic applications covering mid infrared to visible range.~\cite{buscema2014fast}. Field effect transistor have been reported for a few layer BP,~\cite{li2014black,koenig2014electric,liu2014phosphorene,du2014device}
showing on-off ratio of  $\textgreater $10$^5$ as well as high mobility ($\sim$ 1000 cm$^2$/Vs), thereby combining the good aspects of graphene (high mobility) and MoS$_2$ (high on-off ratio).\\
\begin{figure*}[ht!]
 \includegraphics[width=0.8\textwidth]{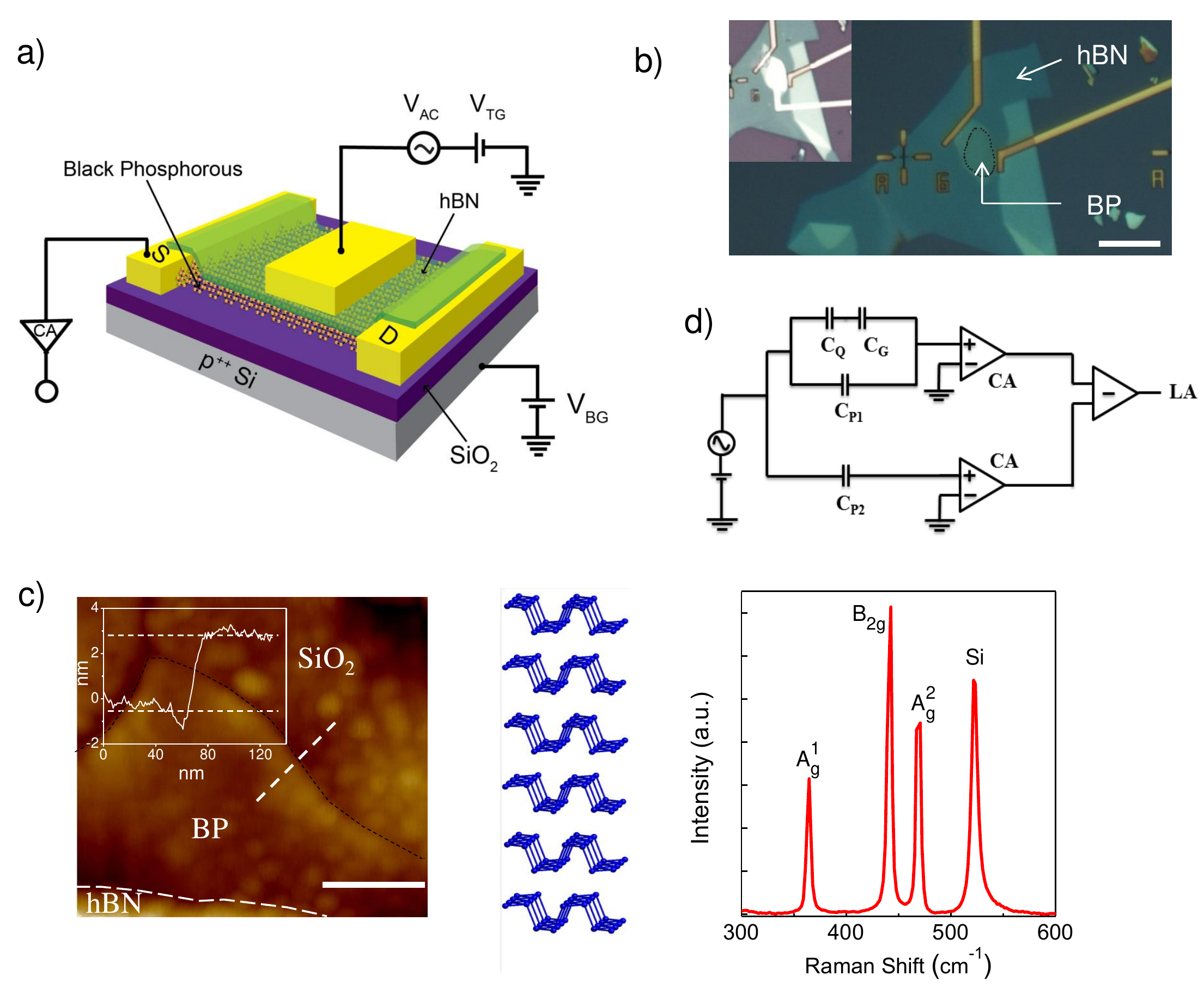}
 \caption{(Color Online) (a) Schematic of the dual gated few layer BP device. (b) Optical image of the device. Scale bar - 10$\mu$m. The black dotted line shows the region of top gate which is shown in the inset. (c) AFM image of the BP (left) with scale bar 100 nm. Inset shows the height profile corresponding to 3 nm BP height (shown by horizontal dashed lines). DFT optimized structure of 6 layer BP (middle). Raman spectra of the BP sample (right). (d) Schematic of the capacitance measurement using differential current amplifier (CA). C$_Q$ + C$_G$ represents the total capacitance between the TG and BP whereas C$_{P1}$ represents parasitic capacitance between the top gate bonding pad and source-drain bonding pad. C$_{P2}$ represents the parasitic capacitance between the top gate bonding pad and a dummy bonding pad (not connected with the device having exact dimension of source and drain bonding pads). LA stands for lock-in amplifier.}
 \label{fig1}
\end{figure*}

In a classical capacitor the capacitance is determined only by the geometry (C$_G$), where the electrons distribute themselves in such a way that the electrostatic energy inversely proportional to the capacitance is minimized. However, addition of an electron requires extra kinetic energy and therefore, contributes to the total capacitance 1/C${_T}$ = 1/C${_G}$ + 1/C${_Q}$, where C$_Q$ is the quantum capacitance.~\cite{luryi1988quantum} The quantum capacitance directly probes the electronic density of states (DOS) of the material.~\cite{ilani2006measurement,dai2009measurements} As a result quantum capacitance can give better insight about the localized states near the band edge of a semiconductor compared to the conductance measurement. Quantum capacitance measurements have been shown to provide significant insights into the ground state of the low dimensional systems such as electron-electron interactions, quantum correlations,~\cite{ilani2006measurement} thermodynamic compressibility for 2D electron gas in GaAs heterostructures~\cite{eisenstein1992negative} and many body physics in carbon nanotubes,~\cite{ilani2006measurement} and graphene.~\cite{xia2009measurement,Yu26022013,chen2008mobility,fang2007carrier,li2013density} Capacitance spectroscopy has been also used to probe metal-insulator transition in MoS$_2$.~\cite{chen2015probing}\\

To investigate the DOS of BP near the band edge, we report here for the first time the quantum capacitance measurements together with the conductance measurements of a few layer BP ($\sim$ 6 layer) at low-temperatures 77 K and 4.2 K as a function of back and top gate voltages. For the later geometry, ultrathin hBN layer ($\sim$ 25 nm thick) has been used as a dielectric which also helps to protect the BP from atmospheric degradation. We compare our quantum capacitance results with the derived density of states based on the density functional theory (DFT). We show that the transport gap from the capacitance measurements is much smaller as compared to that deduced from the conductance measurements. Our result reveal a large asymmetry ($\sim$ 15 times) in the onset voltages between the quantum capacitance and conductance for both the electron and hole sides. These results have been explained in terms of localized states near the band edge as well as asymmetric band dispersion (seen in the DFT) along the zigzag direction ($\Gamma$-Y) for the electron and hole carriers. The presence of localized states is confirmed by the variable range hopping seen in our temperature-dependence conductivity at different values of hole doping achieved by back gate voltage. We have also carried out similar measurements on a few layer MoS$_2$ device and observe a much smaller onset difference between the capacitance and the conductance data.

\section{Results and Discussion}

In order to measure the quantum capacitance of a few layer BP we have used dual gated field effect transistor geometry as shown schematically in Figure 1a. Figure 1b shows the optical image of the device, where a few layer BP mechanically exfoliated from the bulk single crystal (M/s Smart Elements) were transferred onto degenerately doped silicon wafer with 285 nm SiO$_2$ on top. The samples were immediately coated with PMMA layer to prevent degradation of BP. After optical inspection we choose very thin BP flake for the device. The electrical contacts were fabricated using electron beam lithography with Cr/Au (5nm/70nm) contacts deposited using thermal evaporation at a base pressure $\sim$ 3 $\times$ $10^{-7}$ mbar. Using the dry transfer technique described in~\cite{zomer2011transfer}, a thin layer of hBN was transferred immediately after the lift-off on the exposed region of the BP flake, which served as a passivation mask as well as the top gate insulator (inset of Figure 1b). The sample was characterized by atomic force microscope and Raman spectroscopy (Figure 1c). The AFM height measured on our sample is $\sim$ 3 nm corresponding to $\sim$ 6 layers. All the transport measurements were carried out at 4.2 K and 77 K at base pressure of  $\sim$ 1 $\times$ $10^{-6}$ mbar inside a homebuilt dip-stick cryostat. The hBN protected few layer BP was stable without any degradation for more than a month at low temperatures.

\begin{figure*}[ht!]

 \includegraphics[width=0.9\textwidth]{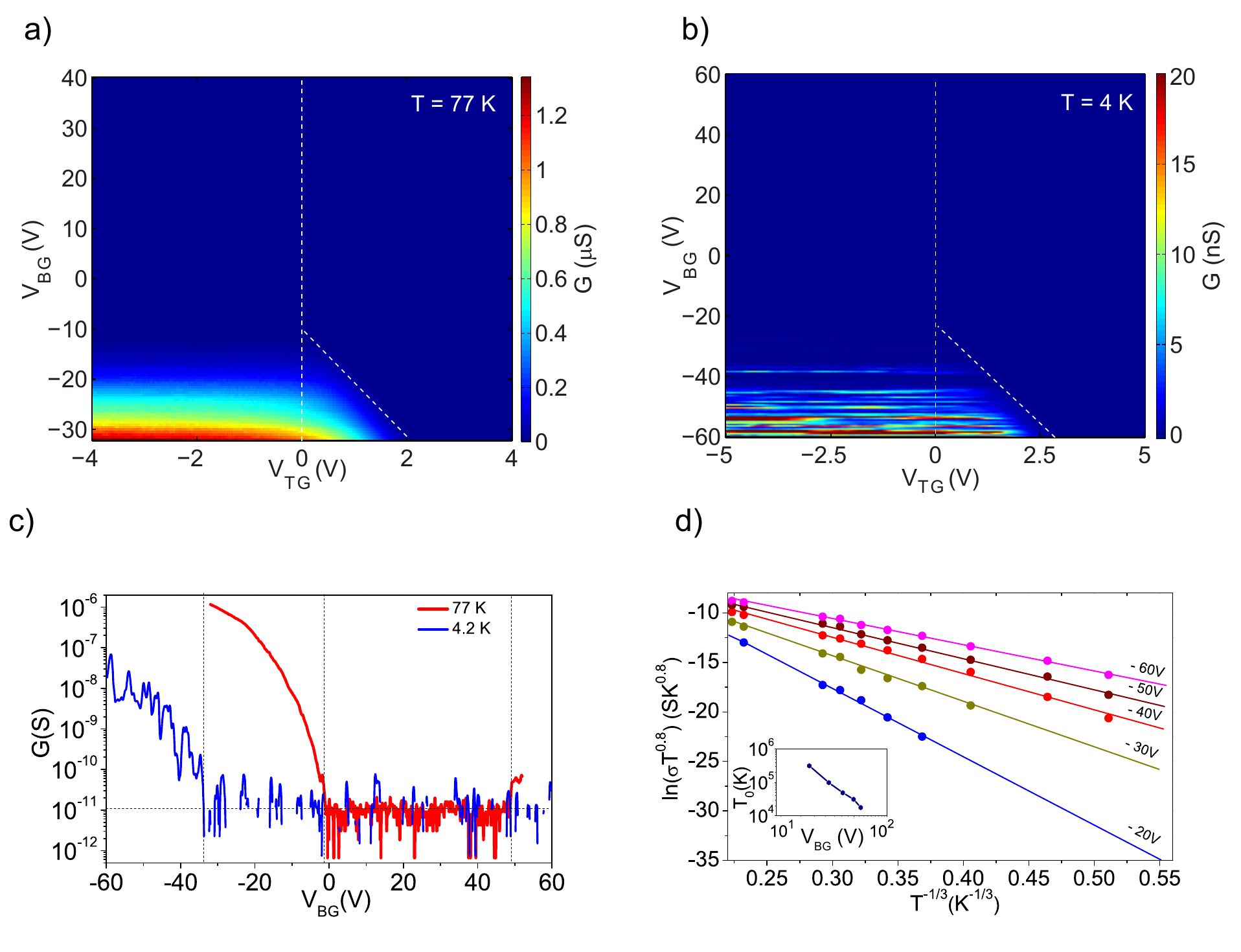}
 \caption{(Color Online) Conductance as a function of V$_{BG}$ and V$_{TG}$ at 77 K (a) and 4.2 K (b) Vertical white dashed lines shows V$_{TG}$ = 0 V cut line. (c) Conductance as a function  of V$_{BG}$ at V$_{TG}$ = 0 V at 77 K and 4.2 K. Vertical black dashed lines show the onset of conductance.(d)  Temperature dependent conductivity and variable range hopping at different back gate voltages. The solid lines are the fit to equation (1). Inset shows T$_0$ for different V$_{BG}$. }
 \label{fig:example}
\end{figure*}

The two probe conductance measurements between the source and drain have been carried out using low frequency ($\sim$ 1 kHz) lock-in technique with a small excitation voltage of 10 mV (at 77 K) and 1 mV (4.2K), using homebuilt current amplifier~\cite{kretinin2012wide} with a gain of 10$^7$. For capacitance measurements we use the circuit diagram as shown in Figure 1d,~\cite{dai2009measurements} using homebuilt differential current amplifier. The capacitance has been measured between the top gate and black phosphorus with a small excitation voltage of 10 mV at $\sim$ 1 kHz frequency with a resolution of $\sim$ 2 fF. The parasitic capacitance (C$_P$) was $\sim$ 100 fF, which reduced to $\sim$ 10 fF after using differential schemes as mentioned above. This value of C$_P$ is subtracted from the measured capacitance data presented in this paper.

As shown in Figure 1a the capacitor is formed between the top gate and the central part of the BP beneath the gate. Figures 2a and 2b show the conductance map ($G=\frac{dI}{dV}$) measured at 77 K and 4.2 K, respectively as a function of top gate (V$_{TG}$) and back gate voltages (V$_{BG}$). The V$_{TG}$ controls the carrier concentration (Fermi energy) in the central part of the BP whereas the back gate controls the carrier concentration throughout the sample, giving \emph{n-p-n} or \emph{p-n-p} like geometries. In principle, an intrinsic semiconductor will conduct whenever the Fermi energy touches the conduction or the valence band edge. In our device, high conductivity is observed only when both sides near the leads and the central part have hole doping (negative gate voltages) showing that BP is unintentionaly \emph{p}-doped at zero gate voltages as known from earlier studies~\cite{li2014black}. From - 5 V to 0 V top gate voltage, the onset of the conductivity starts at a constant back gate voltage as the Fermi energy touches the valence band edge of the leads. However, from 0 V to +2 V of V$_{TG}$ the onset of conductivity depends on both top and back gate voltages. Along the dashed diagonal lines in the two dimensional plot of Figures 2a and 2b , the Fermi energy touches the valence band edge of the central part of the device. From this slope $\frac{dV_{BG}}{dV_{TG}}$ $\sim$ 11 we estimate the thickness of the top hBN to be $\sim$  25 nm ($dV_{BG}$ $\times$ $d_{TG}$ = $dV_{TG}$ $\times$ $d_{BG}$, $d_{BG}$ = 285 nm, where we have taken the dielectric constant  $\varepsilon_{hBN}$ $\sim$ $\varepsilon_{SiO_2}$ = 3.9). In Figure 2c we have shown the conductance in logarithmic scale as a function of V$_{BG}$ at V$_{TG}$ = 0 V (along the vertical dotted line in Figure 2a and 2b) for 77 K and 4.2 K. The on-off ratio in our top gated device is $\sim$ 10$^5$ as can be seen from Figure 2c and field effect mobility measured at 77 K was $\sim$ 20 cm$^2$/Vs  at V$_{BG}$ $\sim$ - 28 V as estimated using the relation $\mu=\frac{L}{WC_G}\frac{dG}{dV_{BG}}$($L = 10 \mu m$, $W = 5 \mu m$ and C$_{G}$ = 1.2 $\times$ 10$^{-8}$ F/cm$^2$). At 77 K we observe the onset of conductance for valence (conduction) bands at - 3 V (+ 50 V), whereas at 4.2 K we only see it for the hole side at $\approx$ - 33 V. The main striking differences at two temperatures are following. With increasing negative back gate voltages the conductance increases smoothly at 77 K whereas at 4.2 K it increases with superimposed random fluctuations and secondly the onset of the conductance with back gate voltage for the hole side increases from 77 K to 4.2 K. These shifts in onset of conductance with lowering temperature can be explained by the mobility edge picture having localized states~\cite{mott2012electronic}. The mobility edge can shift in energy, as the thermally activated transport becomes weaker with lowering the temperature as explained by Mott's Variable Range Hopping (M-VRH) model~\cite{yu2004variable,van1997screening}.
To verify the localization near the band edge we have measured the temperature dependent conductance for hole side as shown in Figure 2d (see supplementary information for the details). The data fits very well with the M-VRH with
\begin{equation}
 \sigma=\sigma_0(T)\exp[-(T_0/T)^{1/(d+1)}]
\end{equation}
,where $\sigma$ is the conductivity, d is the dimensionality of the system, T$_0$ is the correlation energy scale and $\sigma_0$ = AT$^m$ with m $\approx$ 0.8 .The values of T$_0$ are shown in the inset of the Figure 2d.  From the values of T$_0$ (for V$_{BG}$ = - 20 V) and using $\xi^2$ = $\frac{13.8}{k{_B}D(E)T{_0}}$ and taking surface density of electron-hole puddles at SiO$_2$ interface~\cite{ghatak2011nature} $D(E)$ = 4 $\times$ 10$^{12}$ eV$^{-1}$cm$^{-2}$, the localization length $\xi$ is $\approx$ 11 nm, similar to other 2D materials like MoS$_2$~\cite{ghatak2011nature} and WSe$_2$.~\cite{ovchinnikov2014electrical}
These localized states are mostly created due to the charge in homogeneity in the underlying SiO$_2$ layer as well as the defects in BP. Please note that the consideration of contact resistance in our analysis  may change the values of T$_0$. However, it will not alter the main inference drawn as the existence of localized states near the band edge.

\begin{figure*}[ht!]

 \includegraphics[width=0.9\textwidth]{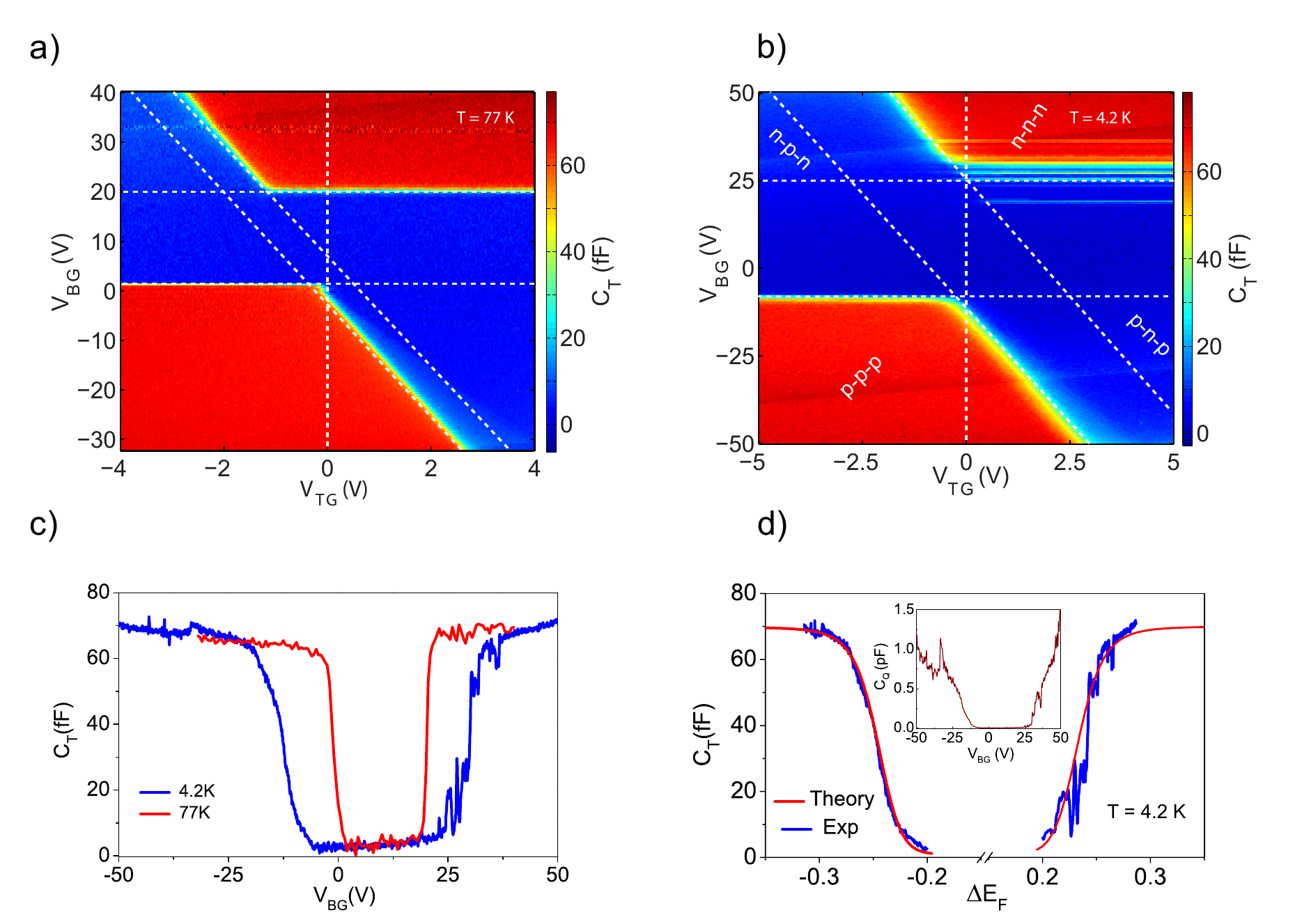}
 \caption{(Color Online) Total capacitance as a function of V$_{BG}$ and V$_{TG}$ at 77 K (a) and 4.2 K (b). (c) Total capacitance as a function of V$_{BG}$ at V$_{TG}$ = 0 V at 77 K (red) and 4.2 K (blue) along the vertical dashed lines in Figure (a) and (b). (d) Comparison between the experimentally measured total capacitance at 4.2 K (blue curve) vs theoretical one (DFT) as a function of Fermi energy (red curve). Inset shows the quantum capacitance as function of back gate voltage.}
 \label{fig:example}
\end{figure*}

We will now present the quantum capacitance measurements. Figure 3a (3b) shows the measured total capacitance as a function of V$_{TG}$ and V$_{BG}$ at 77 K (4.2 K). The four corners of the plots are marked with \emph{p-p-p, p-n-p, n-p-n, n-n-n}, specifying the carrier types near the left side lead, the central part and near the right side lead of the BP. The total capacitance of the central part is zero in the blue region of the 2D plots and it becomes finite, of the order of $\sim$ 70 fF as shown by the red region in the 2D plots for both \emph{p-p-p} and \emph{n-n-n} regions, in contrast with only the hole side conductance (Figure 2a and 2b). The diagonal dashed lines in Figure 3a and 3b indicate the onset of Fermi energy to the band edge in the central part of the BP, whereas the horizontal dashed lines show the onset of Fermi energy to the band edge of the leads for both types of carriers. Even though we are measuring the capacitance of the central part of the BP, we need to tune the leads for \emph{p} or \emph{n} type such that the charging time (RC$_T$) of the central part through the lead resistance (R) becomes smaller than the time period ( $\sim$ 1 ms) of our ac excitation frequency ($\sim$ 1 kHz).For measuring a capacitance of 1 fF in the central part the maximum lead resistance (R) has to be less than 1000 G$\Omega$ .The diagonal dashed lines in Figure 3a and 3b give the hBN thickness  to be $\sim $ 25 nm , in agreement with the conductance measurements. Figure 3c shows the total capacitance as a function of V$_{BG}$ at V$_{TG}$ = 0 V at 77 K and 4.2 K. As mentioned before, 1/C$_T$ = 1/C$_G$ + 1/C$_Q$, therefore, if  C$_Q$ $<<$ C$_G$, C$_Q$ $\sim $ C$_T$. When the Fermi energy is inside the band gap, C$_Q$ will be zero (DOS is zero) and hence C$_T$ will be zero. On the other hand, when the Fermi energy is within the conduction or valence band, C$_Q$ will be much higher than C$_G$ and therefore, C$_T$ $\sim $ C$_G$. At both the temperatures in Figure 3c, the measured C$_Q$ is zero for a certain range of back gate voltages and increases sharply on both sides and saturating to $\sim$ 70 fF. The measured C$_G$ = C$_T$ $\sim$ 70 fF matches very well with the estimated C$_G$ = $\frac{\epsilon A}{d}$  for the gate area (A) of $\sim$ 50 $\mu$m$^2$ (determined from the optical image of the top gate as shown in Figure 1b with d = 25 nm thick hBN). It can be seen from the Figure 3c that the onset of C$_Q$ at 77 K is 0 V (hole side) and + 20 V (electron side) whereas at 4.2 K the onset voltages are - 10 V and + 30 V. The shift in the onset of gate voltages with lowering temperature is symmetric for both types of carriers but there are fluctuations in C$_Q$ (4.2 K) only near the onset on the electron side (Figures 3b and 3c), probably indicating that the localized states near the electron side are more compared to that near the hole side. The extracted C$_Q$ as a function of V$_{BG}$ is shown in the inset of Figure 3d.\\

\begin{figure*}[ht!]

 \includegraphics[width=0.9\textwidth]{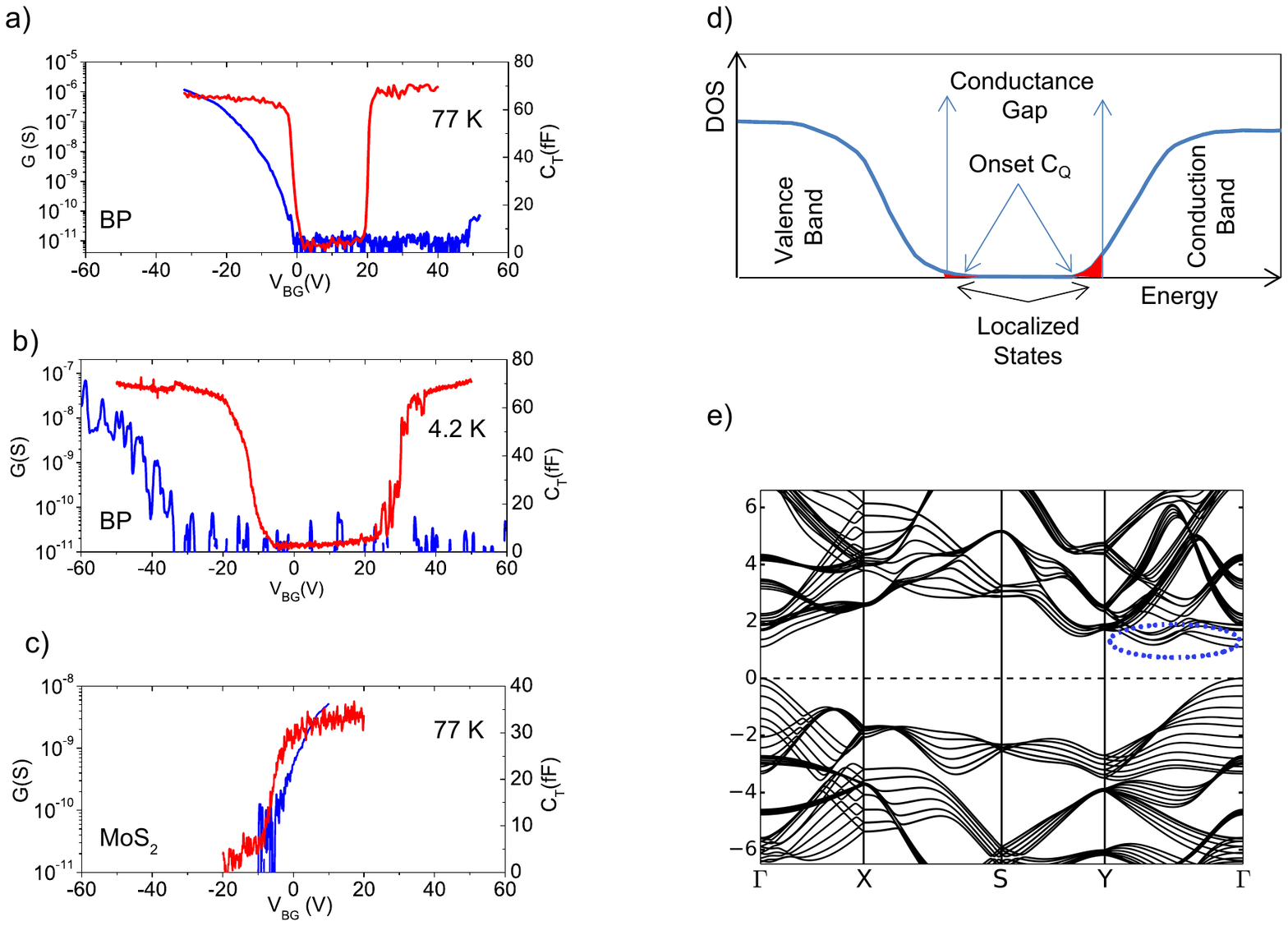}
 \caption{ Onset of conductance (blue) and capacitance (red) as a function of V$_{BG}$ at 77 K (a) and 4.2 K (b) and (c) for few layer MoS$_2$ (d) Schematic of the DOS as a function of energy. (e) Calculated band structure of 6 layer BP. The asymmetry between the electron and holes along $\Gamma$-Y direction is highlighted with blue dotted circle.}
 \label{fig:example}
\end{figure*}

In order to compare the C$_Q$ with the calculated DOS we need to convert the gate voltages into Fermi-energy. E$_F$ = $\alpha$V$_G$ ($\alpha \sim$ is a conversion factor) when the Fermi-energy is inside the band gap and E$_F$ = $f(V_G)$ for the case of Fermi energy in the band. In order to get $\alpha$, we have taken the band gap of BP ($\sim$ 6 layer) to be $\sim$ 400 meV~\cite{li2014black,liu2014phosphorene,das2014tunable} and $\alpha$ becomes 0.013 ($\Delta$V$_G$/$\Delta$E$_F$ = 30/0.4 using the 4.2 K data). Now to determine $f(V_G)$ in the band, we have used the following method. The induced carrier concentration (\emph{n}) in the sample can be determined by
\begin{equation}
n= \frac{1}{e}\int_{V_{G1}}^{V_{G2}} C_{T}(V_{G})\,dV_{G}
\end{equation}
  at different V$_G$ using 4.2 K data from Figure 3c. From the DFT calculation we have calculated the carrier concentration (\emph{n}) as a function of Fermi energy for 6 layer BP. By mapping experimentally determined $n(V_G)$ with theoretical $n(E_F)$, V$_G$ is converted into Fermi-energy (see supplementary information for details). We performed first-principles DFT calculations using the pseudopotential plane wave method as implemented in the Quantum ESPRESSO~\cite{QE-2009} package. A vacuum of 15 {\AA} was employed to avoid spurious interactions between the 6-layer phosphorene and it's periodic images in the out of plane direction. We used norm conserving pseudopotentials~\cite{PhysRevB.43.1993} and accounted for the van der Waals interaction between the layers using the Grimme-D2~\cite{grimme2006semiempirical} formulation. The wavefunctions were expanded in plane waves with energy upto 60 Ry and the Brillouin zone sampled with a 14$\times$10$\times$1 k-point grid. 
The calculations were performed using the Heyd Scuseria Ernzerhof (HSE)~\cite{heyd2003hybrid} as well as the Perdew Burke Ernzerhof (PBE)~\cite{perdew1996rationale} exchange correlation functionals. Within PBE, our calculated structural parameters are in excellent agreement with previous results~\cite{cai2014layer}. However, the calculation within PBE results in a semi-metallic system. This is in contradiction with the experimental result that 6-layer phosphorene is a semiconductor. The band gap of phosphorene is known to reduce with increasing the number of layers~\cite{cai2014layer}. This coupled with the well known understimation of the band gap within Kohn-Sham DFT results in closing of the gap~\cite{PhysRevB.89.235319}. This trend has been reported in the literature and our calculations are in agreement with the previously reported results~\cite{cai2014layer}. While the band gap closing is unphysical, other features of the band structure are unaltered and follow the trends with layer size.To fix the above mentioned problem with the band gap, we performed calculations with the HSE exchange correlation functional. The atom positions here are taken from the PBE relaxed structure. The k-point sampling for the non-local Fock exchange operator is chosen to be 7$\times$5$\times$1 for ease of computation. In order to calculate the density of states on a fine grid of k-points, we transformed the HSE wavefunctions to maximally localized Wannier functions using the Wannier90~\cite{mostofi2008wannier90} package. The DOS was then obtained from a 448$\times$320$\times$1 k-point grid. The full DOS of 6 layer BP has been shown in Figure SF2 in supplementary information. The HSE calculation results in a rigid shift of the PBE valence and conduction bands leading to opening of a band gap. We find the HSE band gap to be about 1 eV, which is larger than the previously reported values. We attribute this discrepancy to the presence of a strain in our structure and the Brillouin zone sampling for the Fock exchange. To compare the theoretical DOS with the experiment we added a Gaussian broadening of 50 meV and rigidly shifted the gap to 400 meV as reported earlier~\cite{li2014black,liu2014phosphorene,das2014tunable}. The theoretical C$_Q$ has been calculated using the relation $e^2 (DOS)$ and in Figure 3d the solid red line shows the theoretical C$_T$ (with C$_G$ = 70 fF), which matches very well with the experimental C$_T$ (solid blue line).\\

Now we will move to the main striking result of this paper. In Figure 4 we have compared the onset of conductance versus the onset of quantum capacitance at 77 K (Figure 4a) and 4.2 K (Figure 4b). At 77 K the differences between the onset voltages $\Delta$V$_{onset}^h$ = V$_{onset}^{conductance}$ - V$_{onset}^{capacitance}$ is $\sim$ 2V for the hole side and $\Delta$V$_{onset}^e$ $\sim$ 30 V for the electron side. At 4.2 K, $\Delta$V$_{onset}^h$ is $\sim$ 20 V whereas we could not measure $\Delta$V$_{onset}^e$, because even upto V$_{BG}$ of 70 V, the device was in the off state in conductance measurement. The quantum capacitance directly measures the DOS and therefore, it can probe the localized states near the band edge as long as the charging takes place. On the other hand the conductance through the channel will occur when the Fermi energy crosses the mobility edge into the extended states, thereby justifying a finite value of $\Delta$V$_{onset}$. However, the large asymmetry between the  $\Delta$V$_{onset}^e$ and $\Delta$V$_{onset}^h$ (R = $\Delta$V$_{onset}^e$/$\Delta$V$_{onset}^h$ $\sim$ 15) is still to be explained. Can this large value of R arise from the different values of mobilities of electrons and holes ($\mu_h \sim$ 20 cm$^2$/Vs and $\mu_e \sim$ 10 cm$^2$/Vs).? To address this point, we did similar measurements on a few layer MoS$_2$ at 77 K having  $\mu_e \sim$ 10cm$^2$/Vs (comparable to that of $\mu_e$(BP)). Figure 4c shows that for MoS$_2$ $\Delta$V$_{onset}^e$ $\sim$ 3 V, which is much smaller than the $\Delta$V$_{onset}^e$ $\sim$ 30 V for the BP. We, therefore infer that the large value of R may not be due to different values of mobilities of electrons and holes. The large value of R suggests that there are more localized states near the conduction band as compared to the valence band, shown schematically in Figure 4d. It is likely that an asymmetry in electronic structure between the electron and hole sides along the ($\Gamma$-Y) direction (highlighted in Figure 4e) can contribute to the above observations. Indeed, Yuan \emph{et.al.} show that the mid gap states, preferentially near the conduction band are created due to point defects in bilayer BP (see Figure 3b in Ref~\cite{yuan2015transport}). However, a quantitative understanding is yet to be emerged for our system.

\section{Conclusion}
In summary, we have measured the quantum capacitance along with the conductance of hBN protected 6 layer BP at low temperatures using both back gate and top gate voltages. The measured capacitance as a function of doping matches with the DFT calculations. The differences in the onset between the conductance and quantum capacitance measurements for the electron and hole sides attributed to the presence of localized states and anisotropic band dispersion, still needs a better theoretical understanding. Our results show that the quantum capacitance can probe the localized states of a two-dimensional semiconductor which cannot be measured by the conductance measurements. It would be worth doing the quantum capacitance measurements of BP as a function of number of layers.\\

\textbf{Acknowledgment}\\

M.J thanks Supercomputing Education and Research Center, IISc for the computational resources. A.K.S thanks the Department of Science and Technology, India under the Nanomission  project for financial support. A.D thanks Department of Atomic Energy and Indian Institute of Science startup grant for financial support.

\section*{Supplementary Information}
The two probe I$_{DS}$ - V$_{DS}$ for different values of V$_{BG}$ is shown in Figure 5a. The linear I-V characterize the ohmic nature of the contacts. The conductance vs V$_{BG}$ at T = 4.2 K is shown in Figure 5b for different excitation voltage. Figure 5c shows the conductance as a function of V$_{BG}$ for different temperatures. The fluctuations in conductance at low temperature are a signature of localized states transport through variable range hopping as mentioned in the main text. The 2D map of differential conductance dI/dV as a function of backgate voltage (V$_{BG}$) and source-drain bias voltage (V$_{DS}$) at 4.2 K is shown in Figure 5d which shows the signature of localized states near the band edge.

\begin{figure}[tbh]
\includegraphics[width=0.5\textwidth]{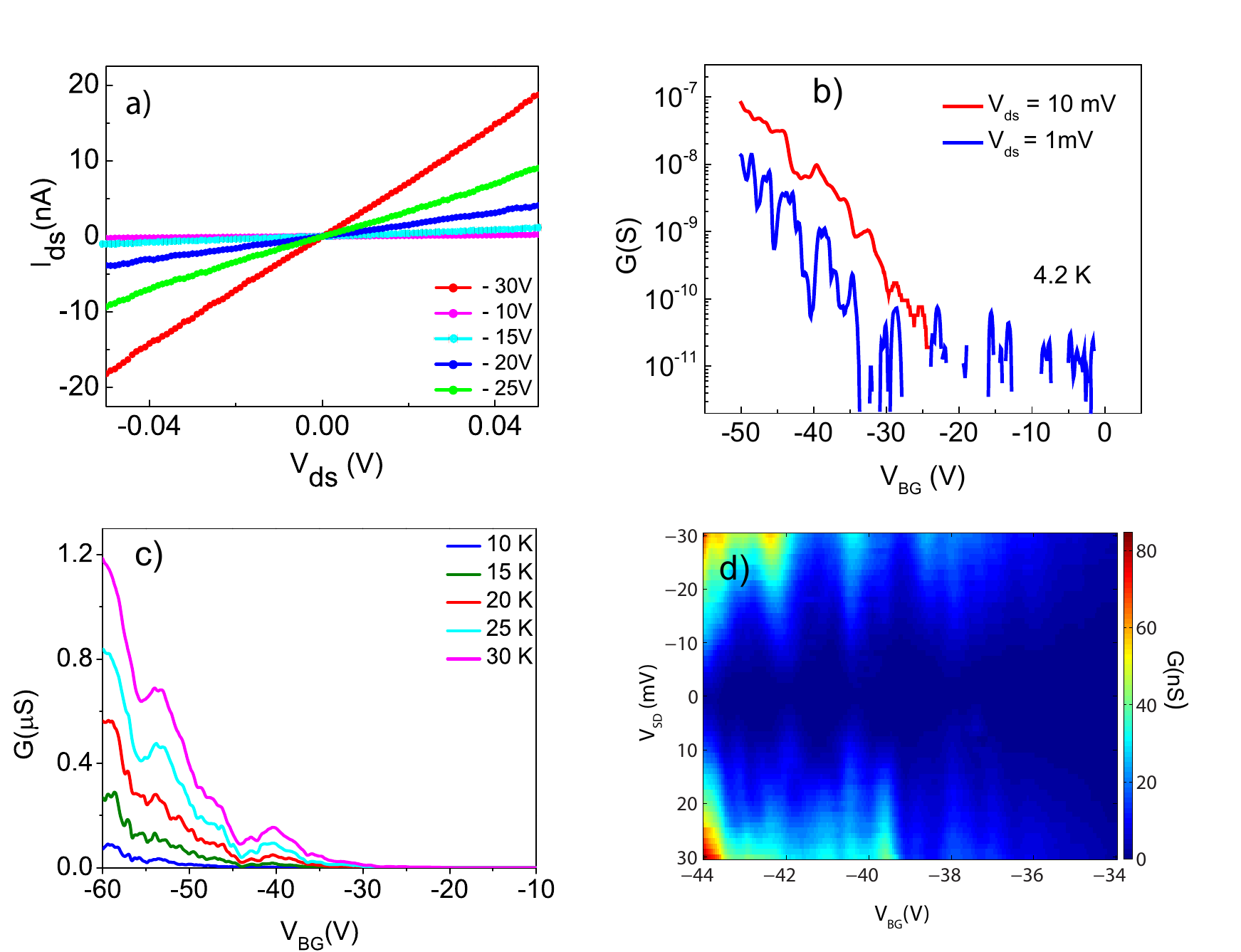}
\small{\caption{ (a) I$_{DS}$ - V$_{DS}$ (b) Conductance vs. V$_{BG}$ at T = 4.2 K for different excitation voltage.(c) Conductance vs. V$_{BG}$ at temperatures 10, 15, 20, 25, and 30 K (d) 2D color map of the differntial conductance as a function  V$_{SD}$ and V$_{BG}$ at T = 4.2 K\label{fig:rt}}}
\end{figure}

The DFT calculations were performed using the Heyd Scuseria Ernzerhof (HSE)~\cite{heyd2003hybrid} as well as the Perdew Burke Ernzerhof (PBE)~\cite{perdew1996rationale} exchange correlation functionals. Within PBE, our calculated structural parameters are in excellent agreement with previous results~\cite{cai2014layer}. However, the calculation within PBE results in a semi-metallic system. This is in contradiction with the experimental result that 6-layer phosphorene is a semiconductor. The band gap of phosphorene is known to reduce with increasing the number of layers~\cite{cai2014layer}. This coupled with the well known understimation of the band gap within Kohn-Sham DFT results in closing of the gap~\cite{tran2014layer}. This trend has been reported in the literature and our calculations are in agreement with the previously reported results~\cite{cai2014layer}. While the band gap closing is unphysical, other features of the band structure are unaltered and follow the trends with layer size.

\begin{figure}[tbh]
\includegraphics[width=0.3\textwidth]{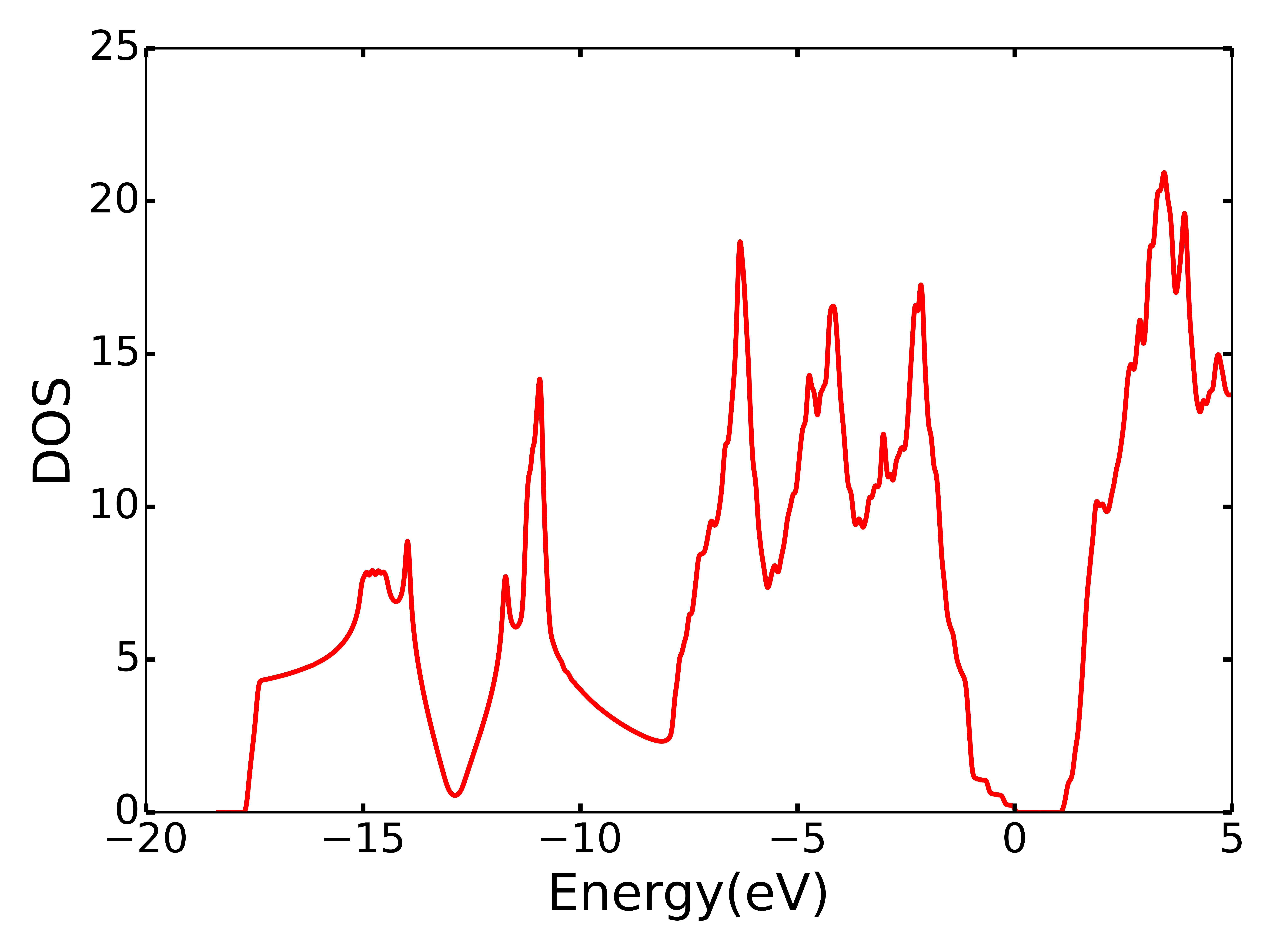}
\small{\caption{Calculated density of states (DOS) vs. Energy \label{fig:rt}}}
\end{figure}

To fix the above mentioned problem with the band gap, we performed calculations with the HSE exchange correlation functional. We used the relaxed structure from PBE for these calculations. The HSE calculation opened a band gap at the $\Gamma$ point which is consistent with experiment. Figure 6 shows the density of states calculated with HSE with a smearing of 50 meV. The band gap ($\sim$ 1eV) we find is larger than that found in previous calculations. We attribute this to the presence of strain in our structure~\cite{morgan2015compressive}. The strain comes about from the use of atom positions relaxed only upto the PBE level. To compare with the experimental data (Figure 3d in the main text) we have taken the band gap $\sim$ 0.4 eV~\cite{li2014black,liu2014phosphorene,das2014tunable} instead of 1 eV.

\begin{figure}[tbh]
\includegraphics[width=0.5\textwidth]{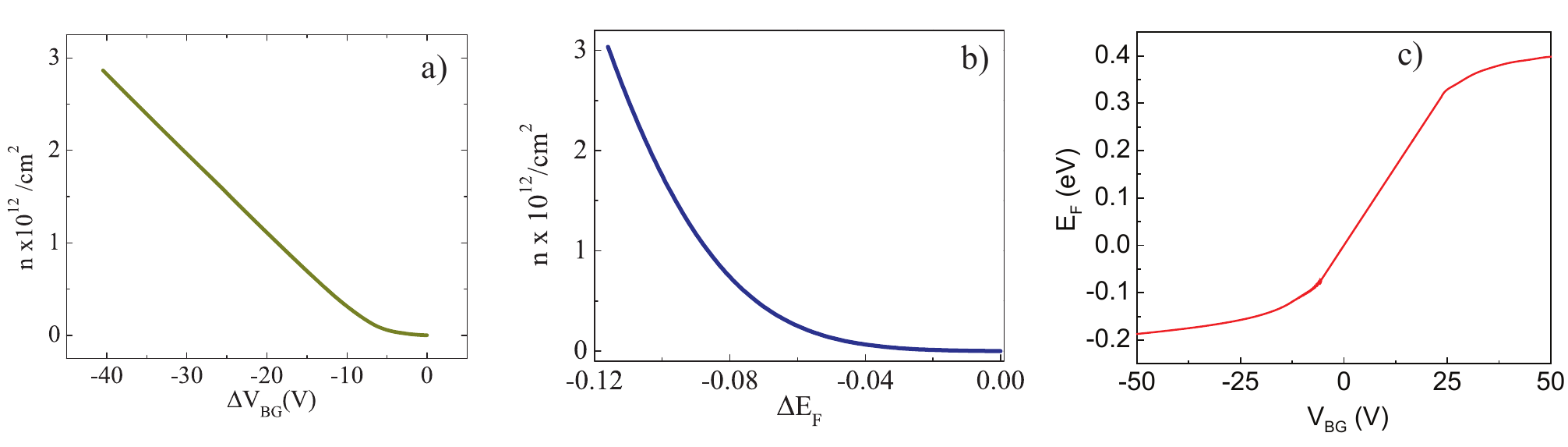}
\small{\caption{(a) $n(V_{BG})$ as a function of $\Delta$V$_{BG}$ for negative gate voltages ( \emph{p} side).
(b) $n(\Delta$E$_F)$ vs $\Delta$E$_F$ obtained from our DFT calculations. (c) Plot of Fermi energy as a function of back gate
voltage.\label{fig:rt}}}
\end{figure}

Effective masses were calculated using the PBE eigenvalues. The electron mass at the $\Gamma$ point, along the armchair direction ($\Gamma$-X), is 0.0267 $m_{el}$ and the hole mass in the same direction is 0.0249 $m_{el}$. The effective mass of the electron and the hole at the $\Gamma$ point, along the zigzag direction ($\Gamma$ - Y), is about 26 times that in the armchair direction. 
This is in good agreement with the trend observed in previous calculations~\cite{cai2014layer}. 

In order to compare the experimental capacitance data with the theoretical one we convert the gate voltage (V$_{G}$) into Fermi energy (E$_F$) in the band. The induced carrier concentration (\emph{n}) in the sample has been determined by

\begin{equation}
n(V_{G})=\frac{1}{e}\int\limits_{V_{G1}}^{V_{G2}}C_{T}(V_{G})dV_{G}
\end{equation}
at different $V_{G}$, where $V_{G1}$ is the onset of capacitance using Figure 3c (4.2 K data) in the main text.

Figure 7a shows the induced carrier density as a function back gate voltage from the onset of capacitance for the hole side. From the DFT calculated DOS we find the carrier concentration as a function of Fermi energy from the onset on hole side which is shown in Figure 7b. By mapping the (\emph{n}) between Figure 7a and 7b, we get the conversion of $\Delta$V$_{BG}$ into $\Delta$E$_F$. Similar analysis was done on the electron side. Figure 7c shows the Fermi energy as a function of backgate voltage.

\section*{References}
\bibliography{references}{}

\begin{thebibliography}{40}%
\makeatletter
\providecommand \@ifxundefined [1]{%
 \@ifx{#1\undefined}
}%
\providecommand \@ifnum [1]{%
 \ifnum #1\expandafter \@firstoftwo
 \else \expandafter \@secondoftwo
 \fi
}%
\providecommand \@ifx [1]{%
 \ifx #1\expandafter \@firstoftwo
 \else \expandafter \@secondoftwo
 \fi
}%
\providecommand \natexlab [1]{#1}%
\providecommand \enquote  [1]{``#1''}%
\providecommand \bibnamefont  [1]{#1}%
\providecommand \bibfnamefont [1]{#1}%
\providecommand \citenamefont [1]{#1}%
\providecommand \href@noop [0]{\@secondoftwo}%
\providecommand \href [0]{\begingroup \@sanitize@url \@href}%
\providecommand \@href[1]{\@@startlink{#1}\@@href}%
\providecommand \@@href[1]{\endgroup#1\@@endlink}%
\providecommand \@sanitize@url [0]{\catcode `\\12\catcode `\$12\catcode
  `\&12\catcode `\#12\catcode `\^12\catcode `\_12\catcode `\%12\relax}%
\providecommand \@@startlink[1]{}%
\providecommand \@@endlink[0]{}%
\providecommand \url  [0]{\begingroup\@sanitize@url \@url }%
\providecommand \@url [1]{\endgroup\@href {#1}{\urlprefix }}%
\providecommand \urlprefix  [0]{URL }%
\providecommand \Eprint [0]{\href }%
\providecommand \doibase [0]{http://dx.doi.org/}%
\providecommand \selectlanguage [0]{\@gobble}%
\providecommand \bibinfo  [0]{\@secondoftwo}%
\providecommand \bibfield  [0]{\@secondoftwo}%
\providecommand \translation [1]{[#1]}%
\providecommand \BibitemOpen [0]{}%
\providecommand \bibitemStop [0]{}%
\providecommand \bibitemNoStop [0]{.\EOS\space}%
\providecommand \EOS [0]{\spacefactor3000\relax}%
\providecommand \BibitemShut  [1]{\csname bibitem#1\endcsname}%
\let\auto@bib@innerbib\@empty
\bibitem [{\citenamefont {Ling}\ \emph {et~al.}(2015)\citenamefont {Ling},
  \citenamefont {Wang}, \citenamefont {Huang}, \citenamefont {Xia},\ and\
  \citenamefont {Dresselhaus}}]{ling2015renaissance}%
  \BibitemOpen
  \bibfield  {author} {\bibinfo {author} {\bibfnamefont {X.}~\bibnamefont
  {Ling}}, \bibinfo {author} {\bibfnamefont {H.}~\bibnamefont {Wang}}, \bibinfo
  {author} {\bibfnamefont {S.}~\bibnamefont {Huang}}, \bibinfo {author}
  {\bibfnamefont {F.}~\bibnamefont {Xia}}, \ and\ \bibinfo {author}
  {\bibfnamefont {M.~S.}\ \bibnamefont {Dresselhaus}},\ }\href@noop {}
  {\bibfield  {journal} {\bibinfo  {journal} {Proc. Natl. Acad. Sci.}\ }\textbf
  {\bibinfo {volume} {112}},\ \bibinfo {pages} {4523} (\bibinfo {year}
  {2015})}\BibitemShut {NoStop}%
\bibitem [{\citenamefont {Li}\ \emph {et~al.}(2014)\citenamefont {Li},
  \citenamefont {Yu}, \citenamefont {Ye}, \citenamefont {Ge}, \citenamefont
  {Ou}, \citenamefont {Wu}, \citenamefont {Feng}, \citenamefont {Chen},\ and\
  \citenamefont {Zhang}}]{li2014black}%
  \BibitemOpen
  \bibfield  {author} {\bibinfo {author} {\bibfnamefont {L.}~\bibnamefont
  {Li}}, \bibinfo {author} {\bibfnamefont {Y.}~\bibnamefont {Yu}}, \bibinfo
  {author} {\bibfnamefont {G.~J.}\ \bibnamefont {Ye}}, \bibinfo {author}
  {\bibfnamefont {Q.}~\bibnamefont {Ge}}, \bibinfo {author} {\bibfnamefont
  {X.}~\bibnamefont {Ou}}, \bibinfo {author} {\bibfnamefont {H.}~\bibnamefont
  {Wu}}, \bibinfo {author} {\bibfnamefont {D.}~\bibnamefont {Feng}}, \bibinfo
  {author} {\bibfnamefont {X.~H.}\ \bibnamefont {Chen}}, \ and\ \bibinfo
  {author} {\bibfnamefont {Y.}~\bibnamefont {Zhang}},\ }\href@noop {}
  {\bibfield  {journal} {\bibinfo  {journal} {Nat. Nanotechnol.}\ }\textbf
  {\bibinfo {volume} {9}},\ \bibinfo {pages} {372} (\bibinfo {year}
  {2014})}\BibitemShut {NoStop}%
\bibitem [{\citenamefont {Tran}\ \emph
  {et~al.}(2014{\natexlab{a}})\citenamefont {Tran}, \citenamefont {Soklaski},
  \citenamefont {Liang},\ and\ \citenamefont {Yang}}]{PhysRevB.89.235319}%
  \BibitemOpen
  \bibfield  {author} {\bibinfo {author} {\bibfnamefont {V.}~\bibnamefont
  {Tran}}, \bibinfo {author} {\bibfnamefont {R.}~\bibnamefont {Soklaski}},
  \bibinfo {author} {\bibfnamefont {Y.}~\bibnamefont {Liang}}, \ and\ \bibinfo
  {author} {\bibfnamefont {L.}~\bibnamefont {Yang}},\ }\href {\doibase
  10.1103/PhysRevB.89.235319} {\bibfield  {journal} {\bibinfo  {journal} {Phys.
  Rev. B}\ }\textbf {\bibinfo {volume} {89}},\ \bibinfo {pages} {235319}
  (\bibinfo {year} {2014}{\natexlab{a}})}\BibitemShut {NoStop}%
\bibitem [{\citenamefont {Qiao}\ \emph {et~al.}(2014)\citenamefont {Qiao},
  \citenamefont {Kong}, \citenamefont {Hu}, \citenamefont {Yang},\ and\
  \citenamefont {Ji}}]{qiao2014high}%
  \BibitemOpen
  \bibfield  {author} {\bibinfo {author} {\bibfnamefont {J.}~\bibnamefont
  {Qiao}}, \bibinfo {author} {\bibfnamefont {X.}~\bibnamefont {Kong}}, \bibinfo
  {author} {\bibfnamefont {Z.-X.}\ \bibnamefont {Hu}}, \bibinfo {author}
  {\bibfnamefont {F.}~\bibnamefont {Yang}}, \ and\ \bibinfo {author}
  {\bibfnamefont {W.}~\bibnamefont {Ji}},\ }\href@noop {} {\bibfield  {journal}
  {\bibinfo  {journal} {Nat. Commun.}\ }\textbf {\bibinfo {volume} {5}}
  (\bibinfo {year} {2014})}\BibitemShut {NoStop}%
\bibitem [{\citenamefont {Wang}\ \emph {et~al.}(2015)\citenamefont {Wang},
  \citenamefont {Jones}, \citenamefont {Seyler}, \citenamefont {Tran},
  \citenamefont {Jia}, \citenamefont {Zhao}, \citenamefont {Wang},
  \citenamefont {Yang}, \citenamefont {Xu},\ and\ \citenamefont
  {Xia}}]{wang2015highly}%
  \BibitemOpen
  \bibfield  {author} {\bibinfo {author} {\bibfnamefont {X.}~\bibnamefont
  {Wang}}, \bibinfo {author} {\bibfnamefont {A.~M.}\ \bibnamefont {Jones}},
  \bibinfo {author} {\bibfnamefont {K.~L.}\ \bibnamefont {Seyler}}, \bibinfo
  {author} {\bibfnamefont {V.}~\bibnamefont {Tran}}, \bibinfo {author}
  {\bibfnamefont {Y.}~\bibnamefont {Jia}}, \bibinfo {author} {\bibfnamefont
  {H.}~\bibnamefont {Zhao}}, \bibinfo {author} {\bibfnamefont {H.}~\bibnamefont
  {Wang}}, \bibinfo {author} {\bibfnamefont {L.}~\bibnamefont {Yang}}, \bibinfo
  {author} {\bibfnamefont {X.}~\bibnamefont {Xu}}, \ and\ \bibinfo {author}
  {\bibfnamefont {F.}~\bibnamefont {Xia}},\ }\href@noop {} {\bibfield
  {journal} {\bibinfo  {journal} {Nature Nanotech.}\ } (\bibinfo {year}
  {2015})}\BibitemShut {NoStop}%
\bibitem [{\citenamefont {Low}\ \emph {et~al.}(2014)\citenamefont {Low},
  \citenamefont {Rodin}, \citenamefont {Carvalho}, \citenamefont {Jiang},
  \citenamefont {Wang}, \citenamefont {Xia},\ and\ \citenamefont
  {Castro~Neto}}]{PhysRevB.90.075434}%
  \BibitemOpen
  \bibfield  {author} {\bibinfo {author} {\bibfnamefont {T.}~\bibnamefont
  {Low}}, \bibinfo {author} {\bibfnamefont {A.~S.}\ \bibnamefont {Rodin}},
  \bibinfo {author} {\bibfnamefont {A.}~\bibnamefont {Carvalho}}, \bibinfo
  {author} {\bibfnamefont {Y.}~\bibnamefont {Jiang}}, \bibinfo {author}
  {\bibfnamefont {H.}~\bibnamefont {Wang}}, \bibinfo {author} {\bibfnamefont
  {F.}~\bibnamefont {Xia}}, \ and\ \bibinfo {author} {\bibfnamefont {A.~H.}\
  \bibnamefont {Castro~Neto}},\ }\href {\doibase 10.1103/PhysRevB.90.075434}
  {\bibfield  {journal} {\bibinfo  {journal} {Phys. Rev. B}\ }\textbf {\bibinfo
  {volume} {90}},\ \bibinfo {pages} {075434} (\bibinfo {year}
  {2014})}\BibitemShut {NoStop}%
\bibitem [{\citenamefont {Castellanos-Gomez}\ \emph {et~al.}(2014)\citenamefont
  {Castellanos-Gomez}, \citenamefont {Vicarelli}, \citenamefont {Prada},
  \citenamefont {Island}, \citenamefont {Narasimha-Acharya}, \citenamefont
  {Blanter}, \citenamefont {Groenendijk}, \citenamefont {Buscema},
  \citenamefont {Steele}, \citenamefont {Alvarez} \emph
  {et~al.}}]{castellanos2014isolation}%
  \BibitemOpen
  \bibfield  {author} {\bibinfo {author} {\bibfnamefont {A.}~\bibnamefont
  {Castellanos-Gomez}}, \bibinfo {author} {\bibfnamefont {L.}~\bibnamefont
  {Vicarelli}}, \bibinfo {author} {\bibfnamefont {E.}~\bibnamefont {Prada}},
  \bibinfo {author} {\bibfnamefont {J.~O.}\ \bibnamefont {Island}}, \bibinfo
  {author} {\bibfnamefont {K.}~\bibnamefont {Narasimha-Acharya}}, \bibinfo
  {author} {\bibfnamefont {S.~I.}\ \bibnamefont {Blanter}}, \bibinfo {author}
  {\bibfnamefont {D.~J.}\ \bibnamefont {Groenendijk}}, \bibinfo {author}
  {\bibfnamefont {M.}~\bibnamefont {Buscema}}, \bibinfo {author} {\bibfnamefont
  {G.~A.}\ \bibnamefont {Steele}}, \bibinfo {author} {\bibfnamefont
  {J.}~\bibnamefont {Alvarez}},  \emph {et~al.},\ }\href@noop {} {\bibfield
  {journal} {\bibinfo  {journal} {2D Materials}\ }\textbf {\bibinfo {volume}
  {1}},\ \bibinfo {pages} {025001} (\bibinfo {year} {2014})}\BibitemShut
  {NoStop}%
\bibitem [{\citenamefont {Das}\ \emph {et~al.}(2014)\citenamefont {Das},
  \citenamefont {Zhang}, \citenamefont {Demarteau}, \citenamefont {Hoffmann},
  \citenamefont {Dubey},\ and\ \citenamefont {Roelofs}}]{das2014tunable}%
  \BibitemOpen
  \bibfield  {author} {\bibinfo {author} {\bibfnamefont {S.}~\bibnamefont
  {Das}}, \bibinfo {author} {\bibfnamefont {W.}~\bibnamefont {Zhang}}, \bibinfo
  {author} {\bibfnamefont {M.}~\bibnamefont {Demarteau}}, \bibinfo {author}
  {\bibfnamefont {A.}~\bibnamefont {Hoffmann}}, \bibinfo {author}
  {\bibfnamefont {M.}~\bibnamefont {Dubey}}, \ and\ \bibinfo {author}
  {\bibfnamefont {A.}~\bibnamefont {Roelofs}},\ }\href@noop {} {\bibfield
  {journal} {\bibinfo  {journal} {Nano Lett.}\ }\textbf {\bibinfo {volume}
  {14}},\ \bibinfo {pages} {5733} (\bibinfo {year} {2014})}\BibitemShut
  {NoStop}%
\bibitem [{\citenamefont {Xia}\ \emph {et~al.}(2014)\citenamefont {Xia},
  \citenamefont {Wang},\ and\ \citenamefont {Jia}}]{xia2014rediscovering}%
  \BibitemOpen
  \bibfield  {author} {\bibinfo {author} {\bibfnamefont {F.}~\bibnamefont
  {Xia}}, \bibinfo {author} {\bibfnamefont {H.}~\bibnamefont {Wang}}, \ and\
  \bibinfo {author} {\bibfnamefont {Y.}~\bibnamefont {Jia}},\ }\href@noop {}
  {\bibfield  {journal} {\bibinfo  {journal} {Nature Comm.}\ }\textbf {\bibinfo
  {volume} {5}} (\bibinfo {year} {2014})}\BibitemShut {NoStop}%
\bibitem [{\citenamefont {Liu}\ \emph {et~al.}(2014)\citenamefont {Liu},
  \citenamefont {Neal}, \citenamefont {Zhu}, \citenamefont {Luo}, \citenamefont
  {Xu}, \citenamefont {Tom{\'a}nek},\ and\ \citenamefont
  {Ye}}]{liu2014phosphorene}%
  \BibitemOpen
  \bibfield  {author} {\bibinfo {author} {\bibfnamefont {H.}~\bibnamefont
  {Liu}}, \bibinfo {author} {\bibfnamefont {A.~T.}\ \bibnamefont {Neal}},
  \bibinfo {author} {\bibfnamefont {Z.}~\bibnamefont {Zhu}}, \bibinfo {author}
  {\bibfnamefont {Z.}~\bibnamefont {Luo}}, \bibinfo {author} {\bibfnamefont
  {X.}~\bibnamefont {Xu}}, \bibinfo {author} {\bibfnamefont {D.}~\bibnamefont
  {Tom{\'a}nek}}, \ and\ \bibinfo {author} {\bibfnamefont {P.~D.}\ \bibnamefont
  {Ye}},\ }\href@noop {} {\bibfield  {journal} {\bibinfo  {journal} {ACS Nano}\
  }\textbf {\bibinfo {volume} {8}},\ \bibinfo {pages} {4033} (\bibinfo {year}
  {2014})}\BibitemShut {NoStop}%
\bibitem [{\citenamefont {Yuan}\ \emph {et~al.}(2015)\citenamefont {Yuan},
  \citenamefont {Rudenko},\ and\ \citenamefont
  {Katsnelson}}]{yuan2015transport}%
  \BibitemOpen
  \bibfield  {author} {\bibinfo {author} {\bibfnamefont {S.}~\bibnamefont
  {Yuan}}, \bibinfo {author} {\bibfnamefont {A.}~\bibnamefont {Rudenko}}, \
  and\ \bibinfo {author} {\bibfnamefont {M.}~\bibnamefont {Katsnelson}},\
  }\href@noop {} {\bibfield  {journal} {\bibinfo  {journal} {Phys. Rev. B}\
  }\textbf {\bibinfo {volume} {91}},\ \bibinfo {pages} {115436} (\bibinfo
  {year} {2015})}\BibitemShut {NoStop}%
\bibitem [{\citenamefont {Buscema}\ \emph {et~al.}(2014)\citenamefont
  {Buscema}, \citenamefont {Groenendijk}, \citenamefont {Blanter},
  \citenamefont {Steele}, \citenamefont {van~der Zant},\ and\ \citenamefont
  {Castellanos-Gomez}}]{buscema2014fast}%
  \BibitemOpen
  \bibfield  {author} {\bibinfo {author} {\bibfnamefont {M.}~\bibnamefont
  {Buscema}}, \bibinfo {author} {\bibfnamefont {D.~J.}\ \bibnamefont
  {Groenendijk}}, \bibinfo {author} {\bibfnamefont {S.~I.}\ \bibnamefont
  {Blanter}}, \bibinfo {author} {\bibfnamefont {G.~A.}\ \bibnamefont {Steele}},
  \bibinfo {author} {\bibfnamefont {H.~S.}\ \bibnamefont {van~der Zant}}, \
  and\ \bibinfo {author} {\bibfnamefont {A.}~\bibnamefont
  {Castellanos-Gomez}},\ }\href@noop {} {\bibfield  {journal} {\bibinfo
  {journal} {Nano Lett.}\ }\textbf {\bibinfo {volume} {14}},\ \bibinfo {pages}
  {3347} (\bibinfo {year} {2014})}\BibitemShut {NoStop}%
\bibitem [{\citenamefont {Koenig}\ \emph {et~al.}(2014)\citenamefont {Koenig},
  \citenamefont {Doganov}, \citenamefont {Schmidt}, \citenamefont {Neto},\ and\
  \citenamefont {Oezyilmaz}}]{koenig2014electric}%
  \BibitemOpen
  \bibfield  {author} {\bibinfo {author} {\bibfnamefont {S.~P.}\ \bibnamefont
  {Koenig}}, \bibinfo {author} {\bibfnamefont {R.~A.}\ \bibnamefont {Doganov}},
  \bibinfo {author} {\bibfnamefont {H.}~\bibnamefont {Schmidt}}, \bibinfo
  {author} {\bibfnamefont {A.~C.}\ \bibnamefont {Neto}}, \ and\ \bibinfo
  {author} {\bibfnamefont {B.}~\bibnamefont {Oezyilmaz}},\ }\href@noop {}
  {\bibfield  {journal} {\bibinfo  {journal} {App. Phys. Lett.}\ }\textbf
  {\bibinfo {volume} {104}},\ \bibinfo {pages} {103106} (\bibinfo {year}
  {2014})}\BibitemShut {NoStop}%
\bibitem [{\citenamefont {Du}\ \emph {et~al.}(2014)\citenamefont {Du},
  \citenamefont {Liu}, \citenamefont {Deng},\ and\ \citenamefont
  {Ye}}]{du2014device}%
  \BibitemOpen
  \bibfield  {author} {\bibinfo {author} {\bibfnamefont {Y.}~\bibnamefont
  {Du}}, \bibinfo {author} {\bibfnamefont {H.}~\bibnamefont {Liu}}, \bibinfo
  {author} {\bibfnamefont {Y.}~\bibnamefont {Deng}}, \ and\ \bibinfo {author}
  {\bibfnamefont {P.~D.}\ \bibnamefont {Ye}},\ }\href@noop {} {\bibfield
  {journal} {\bibinfo  {journal} {ACS Nano}\ }\textbf {\bibinfo {volume} {8}},\
  \bibinfo {pages} {10035} (\bibinfo {year} {2014})}\BibitemShut {NoStop}%
\bibitem [{\citenamefont {Luryi}(1988)}]{luryi1988quantum}%
  \BibitemOpen
  \bibfield  {author} {\bibinfo {author} {\bibfnamefont {S.}~\bibnamefont
  {Luryi}},\ }\href@noop {} {\bibfield  {journal} {\bibinfo  {journal} {App.
  Phys. Lett.}\ }\textbf {\bibinfo {volume} {52}},\ \bibinfo {pages} {501}
  (\bibinfo {year} {1988})}\BibitemShut {NoStop}%
\bibitem [{\citenamefont {Ilani}\ \emph {et~al.}(2006)\citenamefont {Ilani},
  \citenamefont {Donev}, \citenamefont {Kindermann},\ and\ \citenamefont
  {McEuen}}]{ilani2006measurement}%
  \BibitemOpen
  \bibfield  {author} {\bibinfo {author} {\bibfnamefont {S.}~\bibnamefont
  {Ilani}}, \bibinfo {author} {\bibfnamefont {L.~A.}\ \bibnamefont {Donev}},
  \bibinfo {author} {\bibfnamefont {M.}~\bibnamefont {Kindermann}}, \ and\
  \bibinfo {author} {\bibfnamefont {P.~L.}\ \bibnamefont {McEuen}},\
  }\href@noop {} {\bibfield  {journal} {\bibinfo  {journal} {Nat. Phys.}\
  }\textbf {\bibinfo {volume} {2}},\ \bibinfo {pages} {687} (\bibinfo {year}
  {2006})}\BibitemShut {NoStop}%
\bibitem [{\citenamefont {Dai}\ \emph {et~al.}(2009)\citenamefont {Dai},
  \citenamefont {Li}, \citenamefont {Zeng},\ and\ \citenamefont
  {Cui}}]{dai2009measurements}%
  \BibitemOpen
  \bibfield  {author} {\bibinfo {author} {\bibfnamefont {J.}~\bibnamefont
  {Dai}}, \bibinfo {author} {\bibfnamefont {J.}~\bibnamefont {Li}}, \bibinfo
  {author} {\bibfnamefont {H.}~\bibnamefont {Zeng}}, \ and\ \bibinfo {author}
  {\bibfnamefont {X.}~\bibnamefont {Cui}},\ }\href@noop {} {\bibfield
  {journal} {\bibinfo  {journal} {App. Phys. Lett.}\ }\textbf {\bibinfo
  {volume} {94}},\ \bibinfo {pages} {093114} (\bibinfo {year}
  {2009})}\BibitemShut {NoStop}%
\bibitem [{\citenamefont {Eisenstein}\ \emph {et~al.}(1992)\citenamefont
  {Eisenstein}, \citenamefont {Pfeiffer},\ and\ \citenamefont
  {West}}]{eisenstein1992negative}%
  \BibitemOpen
  \bibfield  {author} {\bibinfo {author} {\bibfnamefont {J.}~\bibnamefont
  {Eisenstein}}, \bibinfo {author} {\bibfnamefont {L.}~\bibnamefont
  {Pfeiffer}}, \ and\ \bibinfo {author} {\bibfnamefont {K.}~\bibnamefont
  {West}},\ }\href@noop {} {\bibfield  {journal} {\bibinfo  {journal} {Phys.
  Rev. Lett.}\ }\textbf {\bibinfo {volume} {68}},\ \bibinfo {pages} {674}
  (\bibinfo {year} {1992})}\BibitemShut {NoStop}%
\bibitem [{\citenamefont {Xia}\ \emph {et~al.}(2009)\citenamefont {Xia},
  \citenamefont {Chen}, \citenamefont {Li},\ and\ \citenamefont
  {Tao}}]{xia2009measurement}%
  \BibitemOpen
  \bibfield  {author} {\bibinfo {author} {\bibfnamefont {J.}~\bibnamefont
  {Xia}}, \bibinfo {author} {\bibfnamefont {F.}~\bibnamefont {Chen}}, \bibinfo
  {author} {\bibfnamefont {J.}~\bibnamefont {Li}}, \ and\ \bibinfo {author}
  {\bibfnamefont {N.}~\bibnamefont {Tao}},\ }\href@noop {} {\bibfield
  {journal} {\bibinfo  {journal} {Nat. Nanotech.}\ }\textbf {\bibinfo {volume}
  {4}},\ \bibinfo {pages} {505} (\bibinfo {year} {2009})}\BibitemShut {NoStop}%
\bibitem [{\citenamefont {Yu}\ \emph {et~al.}(2013)\citenamefont {Yu},
  \citenamefont {Jalil}, \citenamefont {Belle}, \citenamefont {Mayorov},
  \citenamefont {Blake}, \citenamefont {Schedin}, \citenamefont {Morozov},
  \citenamefont {Ponomarenko}, \citenamefont {Chiappini}, \citenamefont
  {Wiedmann}, \citenamefont {Zeitler}, \citenamefont {Katsnelson},
  \citenamefont {Geim}, \citenamefont {Novoselov},\ and\ \citenamefont
  {Elias}}]{Yu26022013}%
  \BibitemOpen
  \bibfield  {author} {\bibinfo {author} {\bibfnamefont {G.~L.}\ \bibnamefont
  {Yu}}, \bibinfo {author} {\bibfnamefont {R.}~\bibnamefont {Jalil}}, \bibinfo
  {author} {\bibfnamefont {B.}~\bibnamefont {Belle}}, \bibinfo {author}
  {\bibfnamefont {A.~S.}\ \bibnamefont {Mayorov}}, \bibinfo {author}
  {\bibfnamefont {P.}~\bibnamefont {Blake}}, \bibinfo {author} {\bibfnamefont
  {F.}~\bibnamefont {Schedin}}, \bibinfo {author} {\bibfnamefont {S.~V.}\
  \bibnamefont {Morozov}}, \bibinfo {author} {\bibfnamefont {L.~A.}\
  \bibnamefont {Ponomarenko}}, \bibinfo {author} {\bibfnamefont
  {F.}~\bibnamefont {Chiappini}}, \bibinfo {author} {\bibfnamefont
  {S.}~\bibnamefont {Wiedmann}}, \bibinfo {author} {\bibfnamefont
  {U.}~\bibnamefont {Zeitler}}, \bibinfo {author} {\bibfnamefont {M.~I.}\
  \bibnamefont {Katsnelson}}, \bibinfo {author} {\bibfnamefont {A.~K.}\
  \bibnamefont {Geim}}, \bibinfo {author} {\bibfnamefont {K.~S.}\ \bibnamefont
  {Novoselov}}, \ and\ \bibinfo {author} {\bibfnamefont {D.~C.}\ \bibnamefont
  {Elias}},\ }\href@noop {} {\bibfield  {journal} {\bibinfo  {journal} {Proc.
  Natl. Acad. Sci.}\ }\textbf {\bibinfo {volume} {110}},\ \bibinfo {pages}
  {3282} (\bibinfo {year} {2013})}\BibitemShut {NoStop}%
\bibitem [{\citenamefont {Chen}\ and\ \citenamefont
  {Appenzeller}(2008)}]{chen2008mobility}%
  \BibitemOpen
  \bibfield  {author} {\bibinfo {author} {\bibfnamefont {Z.}~\bibnamefont
  {Chen}}\ and\ \bibinfo {author} {\bibfnamefont {J.}~\bibnamefont
  {Appenzeller}},\ }in\ \href@noop {} {\emph {\bibinfo {booktitle} {Electron
  Devices Meeting, 2008. IEDM 2008. IEEE International}}}\ (\bibinfo
  {organization} {IEEE},\ \bibinfo {year} {2008})\ pp.\ \bibinfo {pages}
  {1--4}\BibitemShut {NoStop}%
\bibitem [{\citenamefont {Fang}\ \emph {et~al.}(2007)\citenamefont {Fang},
  \citenamefont {Konar}, \citenamefont {Xing},\ and\ \citenamefont
  {Jena}}]{fang2007carrier}%
  \BibitemOpen
  \bibfield  {author} {\bibinfo {author} {\bibfnamefont {T.}~\bibnamefont
  {Fang}}, \bibinfo {author} {\bibfnamefont {A.}~\bibnamefont {Konar}},
  \bibinfo {author} {\bibfnamefont {H.}~\bibnamefont {Xing}}, \ and\ \bibinfo
  {author} {\bibfnamefont {D.}~\bibnamefont {Jena}},\ }\href@noop {} {\bibfield
   {journal} {\bibinfo  {journal} {Appl. Phys. Lett.}\ }\textbf {\bibinfo
  {volume} {91}},\ \bibinfo {pages} {092109} (\bibinfo {year}
  {2007})}\BibitemShut {NoStop}%
\bibitem [{\citenamefont {Li}\ \emph {et~al.}(2013)\citenamefont {Li},
  \citenamefont {Chen}, \citenamefont {Wang}, \citenamefont {He}, \citenamefont
  {Wu}, \citenamefont {Cai}, \citenamefont {Zhang}, \citenamefont {Wang},
  \citenamefont {Han}, \citenamefont {Lortz} \emph {et~al.}}]{li2013density}%
  \BibitemOpen
  \bibfield  {author} {\bibinfo {author} {\bibfnamefont {W.}~\bibnamefont
  {Li}}, \bibinfo {author} {\bibfnamefont {X.}~\bibnamefont {Chen}}, \bibinfo
  {author} {\bibfnamefont {L.}~\bibnamefont {Wang}}, \bibinfo {author}
  {\bibfnamefont {Y.}~\bibnamefont {He}}, \bibinfo {author} {\bibfnamefont
  {Z.}~\bibnamefont {Wu}}, \bibinfo {author} {\bibfnamefont {Y.}~\bibnamefont
  {Cai}}, \bibinfo {author} {\bibfnamefont {M.}~\bibnamefont {Zhang}}, \bibinfo
  {author} {\bibfnamefont {Y.}~\bibnamefont {Wang}}, \bibinfo {author}
  {\bibfnamefont {Y.}~\bibnamefont {Han}}, \bibinfo {author} {\bibfnamefont
  {R.~W.}\ \bibnamefont {Lortz}},  \emph {et~al.},\ }\href@noop {} {\bibfield
  {journal} {\bibinfo  {journal} {Scientific Reports}\ }\textbf {\bibinfo
  {volume} {3}} (\bibinfo {year} {2013})}\BibitemShut {NoStop}%
\bibitem [{\citenamefont {Chen}\ \emph {et~al.}(2015)\citenamefont {Chen},
  \citenamefont {Wu}, \citenamefont {Xu}, \citenamefont {Wang}, \citenamefont
  {Huang}, \citenamefont {Han}, \citenamefont {Ye}, \citenamefont {Xiong},
  \citenamefont {Han}, \citenamefont {Long} \emph {et~al.}}]{chen2015probing}%
  \BibitemOpen
  \bibfield  {author} {\bibinfo {author} {\bibfnamefont {X.}~\bibnamefont
  {Chen}}, \bibinfo {author} {\bibfnamefont {Z.}~\bibnamefont {Wu}}, \bibinfo
  {author} {\bibfnamefont {S.}~\bibnamefont {Xu}}, \bibinfo {author}
  {\bibfnamefont {L.}~\bibnamefont {Wang}}, \bibinfo {author} {\bibfnamefont
  {R.}~\bibnamefont {Huang}}, \bibinfo {author} {\bibfnamefont
  {Y.}~\bibnamefont {Han}}, \bibinfo {author} {\bibfnamefont {W.}~\bibnamefont
  {Ye}}, \bibinfo {author} {\bibfnamefont {W.}~\bibnamefont {Xiong}}, \bibinfo
  {author} {\bibfnamefont {T.}~\bibnamefont {Han}}, \bibinfo {author}
  {\bibfnamefont {G.}~\bibnamefont {Long}},  \emph {et~al.},\ }\href@noop {}
  {\bibfield  {journal} {\bibinfo  {journal} {Nature Comm.}\ }\textbf {\bibinfo
  {volume} {6}} (\bibinfo {year} {2015})}\BibitemShut {NoStop}%
\bibitem [{\citenamefont {Zomer}\ \emph {et~al.}(2011)\citenamefont {Zomer},
  \citenamefont {Dash}, \citenamefont {Tombros},\ and\ \citenamefont
  {Van~Wees}}]{zomer2011transfer}%
  \BibitemOpen
  \bibfield  {author} {\bibinfo {author} {\bibfnamefont {P.}~\bibnamefont
  {Zomer}}, \bibinfo {author} {\bibfnamefont {S.}~\bibnamefont {Dash}},
  \bibinfo {author} {\bibfnamefont {N.}~\bibnamefont {Tombros}}, \ and\
  \bibinfo {author} {\bibfnamefont {B.}~\bibnamefont {Van~Wees}},\ }\href@noop
  {} {\bibfield  {journal} {\bibinfo  {journal} {App. Phys. Lett.}\ }\textbf
  {\bibinfo {volume} {99}},\ \bibinfo {pages} {232104} (\bibinfo {year}
  {2011})}\BibitemShut {NoStop}%
\bibitem [{\citenamefont {Kretinin}\ and\ \citenamefont
  {Chung}(2012)}]{kretinin2012wide}%
  \BibitemOpen
  \bibfield  {author} {\bibinfo {author} {\bibfnamefont {A.~V.}\ \bibnamefont
  {Kretinin}}\ and\ \bibinfo {author} {\bibfnamefont {Y.}~\bibnamefont
  {Chung}},\ }\href@noop {} {\bibfield  {journal} {\bibinfo  {journal} {Rev.
  Sci. Instrum.}\ }\textbf {\bibinfo {volume} {83}},\ \bibinfo {pages} {084704}
  (\bibinfo {year} {2012})}\BibitemShut {NoStop}%
\bibitem [{\citenamefont {Mott}\ and\ \citenamefont
  {Davis}(2012)}]{mott2012electronic}%
  \BibitemOpen
  \bibfield  {author} {\bibinfo {author} {\bibfnamefont {N.~F.}\ \bibnamefont
  {Mott}}\ and\ \bibinfo {author} {\bibfnamefont {E.~A.}\ \bibnamefont
  {Davis}},\ }\href@noop {} {\emph {\bibinfo {title} {Electronic Processes in
  Non-Crystalline Materials}}}\ (\bibinfo  {publisher} {Oxford University
  Press},\ \bibinfo {year} {2012})\BibitemShut {NoStop}%
\bibitem [{\citenamefont {Yu}\ \emph {et~al.}(2004)\citenamefont {Yu},
  \citenamefont {Wang}, \citenamefont {Wehrenberg},\ and\ \citenamefont
  {Guyot-Sionnest}}]{yu2004variable}%
  \BibitemOpen
  \bibfield  {author} {\bibinfo {author} {\bibfnamefont {D.}~\bibnamefont
  {Yu}}, \bibinfo {author} {\bibfnamefont {C.}~\bibnamefont {Wang}}, \bibinfo
  {author} {\bibfnamefont {B.~L.}\ \bibnamefont {Wehrenberg}}, \ and\ \bibinfo
  {author} {\bibfnamefont {P.}~\bibnamefont {Guyot-Sionnest}},\ }\href@noop {}
  {\bibfield  {journal} {\bibinfo  {journal} {Phys. Rev. Lett.}\ }\textbf
  {\bibinfo {volume} {92}},\ \bibinfo {pages} {216802} (\bibinfo {year}
  {2004})}\BibitemShut {NoStop}%
\bibitem [{\citenamefont {Van~Keuls}\ \emph {et~al.}(1997)\citenamefont
  {Van~Keuls}, \citenamefont {Hu}, \citenamefont {Jiang},\ and\ \citenamefont
  {Dahm}}]{van1997screening}%
  \BibitemOpen
  \bibfield  {author} {\bibinfo {author} {\bibfnamefont {F.}~\bibnamefont
  {Van~Keuls}}, \bibinfo {author} {\bibfnamefont {X.}~\bibnamefont {Hu}},
  \bibinfo {author} {\bibfnamefont {H.}~\bibnamefont {Jiang}}, \ and\ \bibinfo
  {author} {\bibfnamefont {A.}~\bibnamefont {Dahm}},\ }\href@noop {} {\bibfield
   {journal} {\bibinfo  {journal} {Phys. Rev. B}\ }\textbf {\bibinfo {volume}
  {56}},\ \bibinfo {pages} {1161} (\bibinfo {year} {1997})}\BibitemShut
  {NoStop}%
\bibitem [{\citenamefont {Ghatak}\ \emph {et~al.}(2011)\citenamefont {Ghatak},
  \citenamefont {Pal},\ and\ \citenamefont {Ghosh}}]{ghatak2011nature}%
  \BibitemOpen
  \bibfield  {author} {\bibinfo {author} {\bibfnamefont {S.}~\bibnamefont
  {Ghatak}}, \bibinfo {author} {\bibfnamefont {A.~N.}\ \bibnamefont {Pal}}, \
  and\ \bibinfo {author} {\bibfnamefont {A.}~\bibnamefont {Ghosh}},\
  }\href@noop {} {\bibfield  {journal} {\bibinfo  {journal} {ACS Nano}\
  }\textbf {\bibinfo {volume} {5}},\ \bibinfo {pages} {7707} (\bibinfo {year}
  {2011})}\BibitemShut {NoStop}%
\bibitem [{\citenamefont {Ovchinnikov}\ \emph {et~al.}(2014)\citenamefont
  {Ovchinnikov}, \citenamefont {Allain}, \citenamefont {Huang}, \citenamefont
  {Dumcenco},\ and\ \citenamefont {Kis}}]{ovchinnikov2014electrical}%
  \BibitemOpen
  \bibfield  {author} {\bibinfo {author} {\bibfnamefont {D.}~\bibnamefont
  {Ovchinnikov}}, \bibinfo {author} {\bibfnamefont {A.}~\bibnamefont {Allain}},
  \bibinfo {author} {\bibfnamefont {Y.-S.}\ \bibnamefont {Huang}}, \bibinfo
  {author} {\bibfnamefont {D.}~\bibnamefont {Dumcenco}}, \ and\ \bibinfo
  {author} {\bibfnamefont {A.}~\bibnamefont {Kis}},\ }\href@noop {} {\bibfield
  {journal} {\bibinfo  {journal} {ACS Nano}\ }\textbf {\bibinfo {volume} {8}},\
  \bibinfo {pages} {8174} (\bibinfo {year} {2014})}\BibitemShut {NoStop}%
\bibitem [{\citenamefont {Giannozzi}\ \emph {et~al.}(2009)\citenamefont
  {Giannozzi}, \citenamefont {Baroni}, \citenamefont {Bonini}, \citenamefont
  {Calandra}, \citenamefont {Car}, \citenamefont {Cavazzoni}, \citenamefont
  {Ceresoli}, \citenamefont {Chiarotti}, \citenamefont {Cococcioni},
  \citenamefont {Dabo}, \citenamefont {{Dal Corso}}, \citenamefont
  {de~Gironcoli}, \citenamefont {Fabris}, \citenamefont {Fratesi},
  \citenamefont {Gebauer}, \citenamefont {Gerstmann}, \citenamefont
  {Gougoussis}, \citenamefont {Kokalj}, \citenamefont {Lazzeri}, \citenamefont
  {Martin-Samos}, \citenamefont {Marzari}, \citenamefont {Mauri}, \citenamefont
  {Mazzarello}, \citenamefont {Paolini}, \citenamefont {Pasquarello},
  \citenamefont {Paulatto}, \citenamefont {Sbraccia}, \citenamefont {Scandolo},
  \citenamefont {Sclauzero}, \citenamefont {Seitsonen}, \citenamefont
  {Smogunov}, \citenamefont {Umari},\ and\ \citenamefont
  {Wentzcovitch}}]{QE-2009}%
  \BibitemOpen
  \bibfield  {author} {\bibinfo {author} {\bibfnamefont {P.}~\bibnamefont
  {Giannozzi}}, \bibinfo {author} {\bibfnamefont {S.}~\bibnamefont {Baroni}},
  \bibinfo {author} {\bibfnamefont {N.}~\bibnamefont {Bonini}}, \bibinfo
  {author} {\bibfnamefont {M.}~\bibnamefont {Calandra}}, \bibinfo {author}
  {\bibfnamefont {R.}~\bibnamefont {Car}}, \bibinfo {author} {\bibfnamefont
  {C.}~\bibnamefont {Cavazzoni}}, \bibinfo {author} {\bibfnamefont
  {D.}~\bibnamefont {Ceresoli}}, \bibinfo {author} {\bibfnamefont {G.~L.}\
  \bibnamefont {Chiarotti}}, \bibinfo {author} {\bibfnamefont {M.}~\bibnamefont
  {Cococcioni}}, \bibinfo {author} {\bibfnamefont {I.}~\bibnamefont {Dabo}},
  \bibinfo {author} {\bibfnamefont {A.}~\bibnamefont {{Dal Corso}}}, \bibinfo
  {author} {\bibfnamefont {S.}~\bibnamefont {de~Gironcoli}}, \bibinfo {author}
  {\bibfnamefont {S.}~\bibnamefont {Fabris}}, \bibinfo {author} {\bibfnamefont
  {G.}~\bibnamefont {Fratesi}}, \bibinfo {author} {\bibfnamefont
  {R.}~\bibnamefont {Gebauer}}, \bibinfo {author} {\bibfnamefont
  {U.}~\bibnamefont {Gerstmann}}, \bibinfo {author} {\bibfnamefont
  {C.}~\bibnamefont {Gougoussis}}, \bibinfo {author} {\bibfnamefont
  {A.}~\bibnamefont {Kokalj}}, \bibinfo {author} {\bibfnamefont
  {M.}~\bibnamefont {Lazzeri}}, \bibinfo {author} {\bibfnamefont
  {L.}~\bibnamefont {Martin-Samos}}, \bibinfo {author} {\bibfnamefont
  {N.}~\bibnamefont {Marzari}}, \bibinfo {author} {\bibfnamefont
  {F.}~\bibnamefont {Mauri}}, \bibinfo {author} {\bibfnamefont
  {R.}~\bibnamefont {Mazzarello}}, \bibinfo {author} {\bibfnamefont
  {S.}~\bibnamefont {Paolini}}, \bibinfo {author} {\bibfnamefont
  {A.}~\bibnamefont {Pasquarello}}, \bibinfo {author} {\bibfnamefont
  {L.}~\bibnamefont {Paulatto}}, \bibinfo {author} {\bibfnamefont
  {C.}~\bibnamefont {Sbraccia}}, \bibinfo {author} {\bibfnamefont
  {S.}~\bibnamefont {Scandolo}}, \bibinfo {author} {\bibfnamefont
  {G.}~\bibnamefont {Sclauzero}}, \bibinfo {author} {\bibfnamefont {A.~P.}\
  \bibnamefont {Seitsonen}}, \bibinfo {author} {\bibfnamefont {A.}~\bibnamefont
  {Smogunov}}, \bibinfo {author} {\bibfnamefont {P.}~\bibnamefont {Umari}}, \
  and\ \bibinfo {author} {\bibfnamefont {R.~M.}\ \bibnamefont {Wentzcovitch}},\
  }\href {http://www.quantum-espresso.org} {\bibfield  {journal} {\bibinfo
  {journal} {Journal of Physics: Condensed Matter}\ }\textbf {\bibinfo {volume}
  {21}},\ \bibinfo {pages} {395502} (\bibinfo {year} {2009})}\BibitemShut
  {NoStop}%
\bibitem [{\citenamefont {Troullier}\ and\ \citenamefont
  {Martins}(1991)}]{PhysRevB.43.1993}%
  \BibitemOpen
  \bibfield  {author} {\bibinfo {author} {\bibfnamefont {N.}~\bibnamefont
  {Troullier}}\ and\ \bibinfo {author} {\bibfnamefont {J.~L.}\ \bibnamefont
  {Martins}},\ }\href {\doibase 10.1103/PhysRevB.43.1993} {\bibfield  {journal}
  {\bibinfo  {journal} {Phys. Rev. B}\ }\textbf {\bibinfo {volume} {43}},\
  \bibinfo {pages} {1993} (\bibinfo {year} {1991})}\BibitemShut {NoStop}%
\bibitem [{\citenamefont {Grimme}(2006)}]{grimme2006semiempirical}%
  \BibitemOpen
  \bibfield  {author} {\bibinfo {author} {\bibfnamefont {S.}~\bibnamefont
  {Grimme}},\ }\href@noop {} {\bibfield  {journal} {\bibinfo  {journal}
  {Journal of Computational Chemistry}\ }\textbf {\bibinfo {volume} {27}},\
  \bibinfo {pages} {1787} (\bibinfo {year} {2006})}\BibitemShut {NoStop}%
\bibitem [{\citenamefont {Heyd}\ \emph {et~al.}(2003)\citenamefont {Heyd},
  \citenamefont {Scuseria},\ and\ \citenamefont {Ernzerhof}}]{heyd2003hybrid}%
  \BibitemOpen
  \bibfield  {author} {\bibinfo {author} {\bibfnamefont {J.}~\bibnamefont
  {Heyd}}, \bibinfo {author} {\bibfnamefont {G.~E.}\ \bibnamefont {Scuseria}},
  \ and\ \bibinfo {author} {\bibfnamefont {M.}~\bibnamefont {Ernzerhof}},\
  }\href@noop {} {\bibfield  {journal} {\bibinfo  {journal} {J. Chem. Phys.}\
  }\textbf {\bibinfo {volume} {118}},\ \bibinfo {pages} {8207} (\bibinfo {year}
  {2003})}\BibitemShut {NoStop}%
\bibitem [{\citenamefont {Perdew}\ \emph {et~al.}(1996)\citenamefont {Perdew},
  \citenamefont {Ernzerhof},\ and\ \citenamefont
  {Burke}}]{perdew1996rationale}%
  \BibitemOpen
  \bibfield  {author} {\bibinfo {author} {\bibfnamefont {J.~P.}\ \bibnamefont
  {Perdew}}, \bibinfo {author} {\bibfnamefont {M.}~\bibnamefont {Ernzerhof}}, \
  and\ \bibinfo {author} {\bibfnamefont {K.}~\bibnamefont {Burke}},\
  }\href@noop {} {\bibfield  {journal} {\bibinfo  {journal} {J. Chem. Phys.}\
  }\textbf {\bibinfo {volume} {105}},\ \bibinfo {pages} {9982} (\bibinfo {year}
  {1996})}\BibitemShut {NoStop}%
\bibitem [{\citenamefont {Cai}\ \emph {et~al.}(2014)\citenamefont {Cai},
  \citenamefont {Zhang},\ and\ \citenamefont {Zhang}}]{cai2014layer}%
  \BibitemOpen
  \bibfield  {author} {\bibinfo {author} {\bibfnamefont {Y.}~\bibnamefont
  {Cai}}, \bibinfo {author} {\bibfnamefont {G.}~\bibnamefont {Zhang}}, \ and\
  \bibinfo {author} {\bibfnamefont {Y.-W.}\ \bibnamefont {Zhang}},\ }\href@noop
  {} {\bibfield  {journal} {\bibinfo  {journal} {Scientific Reports}\ }\textbf
  {\bibinfo {volume} {4}} (\bibinfo {year} {2014})}\BibitemShut {NoStop}%
\bibitem [{\citenamefont {Mostofi}\ \emph {et~al.}(2008)\citenamefont
  {Mostofi}, \citenamefont {Yates}, \citenamefont {Lee}, \citenamefont {Souza},
  \citenamefont {Vanderbilt},\ and\ \citenamefont
  {Marzari}}]{mostofi2008wannier90}%
  \BibitemOpen
  \bibfield  {author} {\bibinfo {author} {\bibfnamefont {A.~A.}\ \bibnamefont
  {Mostofi}}, \bibinfo {author} {\bibfnamefont {J.~R.}\ \bibnamefont {Yates}},
  \bibinfo {author} {\bibfnamefont {Y.-S.}\ \bibnamefont {Lee}}, \bibinfo
  {author} {\bibfnamefont {I.}~\bibnamefont {Souza}}, \bibinfo {author}
  {\bibfnamefont {D.}~\bibnamefont {Vanderbilt}}, \ and\ \bibinfo {author}
  {\bibfnamefont {N.}~\bibnamefont {Marzari}},\ }\href@noop {} {\bibfield
  {journal} {\bibinfo  {journal} {Computer physics communications}\ }\textbf
  {\bibinfo {volume} {178}},\ \bibinfo {pages} {685} (\bibinfo {year}
  {2008})}\BibitemShut {NoStop}%
\bibitem [{\citenamefont {Tran}\ \emph
  {et~al.}(2014{\natexlab{b}})\citenamefont {Tran}, \citenamefont {Soklaski},
  \citenamefont {Liang},\ and\ \citenamefont {Yang}}]{tran2014layer}%
  \BibitemOpen
  \bibfield  {author} {\bibinfo {author} {\bibfnamefont {V.}~\bibnamefont
  {Tran}}, \bibinfo {author} {\bibfnamefont {R.}~\bibnamefont {Soklaski}},
  \bibinfo {author} {\bibfnamefont {Y.}~\bibnamefont {Liang}}, \ and\ \bibinfo
  {author} {\bibfnamefont {L.}~\bibnamefont {Yang}},\ }\href@noop {} {\bibfield
   {journal} {\bibinfo  {journal} {Phys. Rev. B}\ }\textbf {\bibinfo {volume}
  {89}},\ \bibinfo {pages} {235319} (\bibinfo {year}
  {2014}{\natexlab{b}})}\BibitemShut {NoStop}%
\bibitem [{\citenamefont {Morgan~Stewart}\ \emph {et~al.}(2015)\citenamefont
  {Morgan~Stewart}, \citenamefont {Shevlin}, \citenamefont {Catlow},\ and\
  \citenamefont {Guo}}]{morgan2015compressive}%
  \BibitemOpen
  \bibfield  {author} {\bibinfo {author} {\bibfnamefont {H.}~\bibnamefont
  {Morgan~Stewart}}, \bibinfo {author} {\bibfnamefont {S.~A.}\ \bibnamefont
  {Shevlin}}, \bibinfo {author} {\bibfnamefont {C.~R.~A.}\ \bibnamefont
  {Catlow}}, \ and\ \bibinfo {author} {\bibfnamefont {Z.~X.}\ \bibnamefont
  {Guo}},\ }\href@noop {} {\bibfield  {journal} {\bibinfo  {journal} {Nano
  Lett.}\ }\textbf {\bibinfo {volume} {15}},\ \bibinfo {pages} {2006} (\bibinfo
  {year} {2015})}\BibitemShut {NoStop}%
\end{thebibliography}%


%

\end{document}